\documentstyle[aps,pre,psfig]{revtex}

\newcommand{\beq}{\begin{equation}}
\newcommand{\eeq}{\end{equation}}
\newcommand{\bea}{\begin{eqnarray}}
\newcommand{\eea}{\end{eqnarray}}

\begin{document}

\draft

\title{\bf Hamiltonian dynamics of the two-dimensional lattice $\varphi^4$
model}

\author{Lando Caiani\cite{lando}}
\address{Scuola Internazionale Superiore di Studi Avanzati (SISSA/ISAS), 
via Beirut 2-4, I-10134 Trieste, Italy} 

\author{Lapo Casetti\cite{lapo}}
\address{Istituto Nazionale di Fisica della Materia (INFM), Unit\`a di Ricerca
del Politecnico di Torino, \\
Dipartimento di Fisica, Politecnico di Torino, 
Corso Duca degli Abruzzi 24, I-10129 Torino, Italy} 

\author{Marco Pettini\cite{marco}}
\address{Osservatorio Astrofisico di Arcetri, Largo Enrico Fermi 5,
I-50125 Firenze, Italy} 

\date {\today}
\maketitle

\begin{abstract}
The Hamiltonian dynamics of the classical $\varphi^4$ model
on a two-dimensional square lattice is investigated by means
of numerical simulations. The macroscopic observables are
computed as time averages. The results clearly reveal the 
presence of the continuous phase transition at a finite energy
density and are consistent both qualitatively and quantitatively
with the predictions of equilibrium statistical mechanics.
The Hamiltonian microscopic dynamics also exhibits critical slowing
down close to the transition. Moreover, the relationship between
chaos and the phase transition is considered, and interpreted in
the light of a geometrization of dynamics. 
\end{abstract}

\section{Introduction}

The study of the interplay between microscopic, deterministic 
dynamics and macroscopic statistical
behaviour of large Hamiltonian systems is an old subject which dates back
to Boltzmann. This subject has been mainly approached having in
mind the problem of the dynamical foundations of equilibrium 
statistical mechanics. In this framework the so-called ergodic
problem is the central point: dynamics is studied in the
perspective of proving ergodicity and mixing, thus giving a
sound foundation to equilibrium statistical mechanics. 
Despite many efforts in ergodic theory, this goal remains
distant, ergodicity and mixing having been proved only for abstract
systems like the Sinai billiard.

However, a 
different approach is possible: instead of making a
statistical assumption and then looking for its
justification in the properties of the microscopic dynamics,
one can consider dynamics from the very beginning. In
practice, instead of considering a particular Hamiltonian
system and trying to prove that its dynamics is mixing, one
can observe the actual dynamical evolution of the system and
measure the time averages of the dynamical observables of
interest. In this way the statistical behaviour emerges
directly from the dynamics, and one can wonder whether this
behaviour is consistent with the predictions of statistical
mechanics.
In general, such an approach needs a tool that was not
available at Boltzmann's times, i.e., a fast computer. In
fact, this approach was pioneered by Fermi, Pasta, and Ulam \cite{FPU},
who performed, around 1950, the first numerical experiment
on the relationship between dynamics and statistical mechanics,
using one of the first computing machines, the
MANIAC computer at Los Alamos. 

The conceptual point of view adopted in the present work is the
Fermi, Pasta, and Ulam's one. We consider the dynamics from
the very beginning, and we look at the statistical
properties as emerging from the dynamics itself. The main goal
of the present work is to show that
a dynamical approach is worth and can provide some genuinely
new understanding of phenomena that are usually treated in
the framework of equilibrium statistical mechanics, like
phase transitions. We argue that some of the information
that is present in the dynamics of a Hamiltonian system, and
that is thrown away at the beginning of the
statistical-mechanical description, is relevant to the
cooperative phenomena that show up in connection with the
phase transition. This means that the relevance of dynamics
in this context might go well beyond the foundational
aspects related with the ergodic problem.

In particular, by
studying the dynamics of a paradigmatic system (belonging to
the universality class of the two-dimensional Ising model), the
$\varphi^4$ model on a two-dimensional lattice, we are able
to show that the dynamical description not only is
consistent with the statistical-mechanical results, but also that
a Hamiltonian description of dynamical phenomena like the
critical slowing down is possible. Moreover, we suggest that
a geometrization of Hamiltonian dynamics based on simple
tools of Riemannian geometry --- originally introduced to
describe chaotic dynamics 
\cite{others,Pettini,CasettiPettini,CerrutiPettini,PettiniValdettaro,prl95,CCP}
--- can provide a global description of the dynamical properties that are
relevant to the statistical behaviour. We argue that most of
these properties are indeed the consequence of the actual
geometric structure of the manifolds where the motion takes
place. Hence geometry can be not only a useful tool in the
theory of chaos, but hopefully can also provide the right
language to bridge microscopic dynamics
and macroscopic statistical behaviour which is still
lacking. In particular, phase transitions might be seen
as the consequence of some major change in the geometric
or even in the topologic structure of the ``mechanical''
manifolds \cite{cccp}. Thus dynamics can bring in new concepts and
tools in statistical physics. In our opinion these tools ---
here applied to a ``standard'' statistical-mechanical system ---
might prove useful both on conceptual and on practical grounds
also dealing with ``non-standard'' topics in statistical physics,
as the emerging field of phase transitions in finite systems
(clusters, polymers, proteins) or long-studied but still unsolved problems
like the dynamics and the statistical mechanics of glasses and
more generally disordered or frustrated systems.

Several other works have recently addressed the problem of
the relevance of microscopic Hamiltonian dynamics to phase transitions,
in particular in the framework of mean-field-like models
\cite{Ruffo} and more generally as far as long-range couplings 
among particles are considered \cite{Antoni}. Moreover, there is
now a renewed interest in microcanonical thermodynamics, 
both on general aspects \cite{Rugh,Giardina} and on phase transitions 
\cite{Gross,Dellago}. 

The paper is organized as follows: in Section \ref{sec_general} 
we discuss some general aspects of the problem, then in Section 
\ref{sec_model} the model and the relevant dynamical observables 
are introduced. Section \ref{sec_results} is devoted to a discussion
of the results of the dynamical simulations. In Section \ref{sec_geometry}
we briefly recall the main points of the Riemannian theory of 
Hamiltonian dynamics and then we discuss the peculiar geometric
properties of the two-dimensional $\varphi^4$ lattice model. 
Finally in Section \ref{sec_conclusions} we draw some conclusions.

\section{Hamiltonian dynamics and phase transitions}

\label{sec_general}

The dynamical aspects of equilibrium phase transitions are 
usually approached assuming from the 
outset the formalism of canonical equilibrium statistical 
mechanics. Dynamics is introduced only {\em a 
posteriori}, and usually making use of phenomenological, 
non-deterministic dynamics, e.g. Langevin dynamics
\cite{HohenbergHalperin}.

As already stated in the Introduction, our approach is different.
Given a Hamiltonian system which
exhibits a phase transition according to equilibrium 
statistical mechanics,
we wonder what is its dynamical 
behaviour when it is studied as a natural dynamical system, 
i.e., associating {\em a priori} to the system a 
deterministic (Hamiltonian) dynamics. Accordingly, thermodynamic 
quantities will be obtained as time averages 
along the dynamical trajectories, whence --- assuming 
ergodicity at least for the ``usual'' thermodynamic 
observables (see Ref. \cite{Giardina} for recent results 
on the dependence of ergodic behaviour on the choice of the observable)
--- such quantities will be equal 
to {\em microcanonical} averages, whereas second-order equilibrium 
phase transitions are usually studied within the {\em canonical} 
ensemble. 

The two ensembles are
equivalent only in the thermodynamic limit, thus the phenomenology observed
in finite systems, as systems considered in numerical simulations 
necessarily are, might be different. To give only one example, let us 
consider the phenomenon of {\em ergodicity breaking}, i.e., the fact that
ergodicity might no longer hold in the whole phase space but only
in disjoint subsets of it. Such a phenomenon is indeed 
tightly related to phase
transitions; in fact, when it happens, one may observe, as a consequence,
a symmetry breaking, as in usual phase transitions. But
ergodicity breaking is a more general concept than symmetry breaking,
in fact one can recognize ergodicity breaking also as the origin of
those phase transitions which do not correspond to the breaking of an
evident symmetry of the Hamiltonian (for example in spin glasses) 
\cite{Goldenfeld}.
In the canonical ensemble, ergodicity can be broken 
only in the thermodynamic
limit, while in the microcanonical ensemble, in principle,
there might be ergodicity breaking also in finite systems. Ergodicity 
being a dynamical property, we think that a dynamical
approach is particularly appropriate to study such a phenomenon.

In the following we shall present in detail a dynamical 
study, pefrormed by means of numerical simulations, of the 
so-called $\varphi^4$ lattice model with ${\bf Z}_2$ 
symmetry. The first, preliminary, goal of such a study is to 
show that the dynamical phenomenology is consistent with the 
usual statistical description (the model exhibits a 
continuous phase transition which belongs to the 
universality class of the Ising model). 
The main results concern intrinsically dynamical 
properties, i.e., time correlation functions 
and Lyapunov exponents, and suggest  
interesting developments. In particular, two facts emerge:
$(i)$ the possibility of a
Hamiltonian (i.e., {\em ab initio}) 
description of critical dynamics aspects like the critical slowing down, 
and $(ii)$ a tight relationship between
the {\em local} instability of the dynamics in phase space 
(characterized by the Lyapunov exponent and the related
geometric observables discussed in Sec. \ref{sec_geometry}) on the
one side and the {\em global} phenomenon of the phase transition
on the other side.
It is worth noticing
that a peculiar behaviour of the temperature-dependence of the
Lyapunov exponent close to a phase transition (actually a
Kosterlitz-Thouless transition) was observed for the first time
by Butera and Caravati more than ten years ago \cite{ButeraCaravati}.

A more detailed discussion of the above-mentioned
aspects as well as of the results can be found in Ref. \cite{thesis}.

\section{Model and dynamical observables}

\label{sec_model}

Let us consider a discretized (lattice) version of 
the classical $\varphi^4$ field
Hamiltonian, which reads as
\beq
{\cal H} =
\sum_i \left[\frac{1}{2}\pi_{i}^2 + 
\frac{J}{2}
\sum_{j=1}^d
(\varphi_{{i}+{j}}-\varphi_{i})^2-\frac{1}{2} 
m^2\varphi_{i}^2
+\frac{\lambda}{4!}\varphi_{i}^4\right]~.
\label{hfi4d}
\eeq

According to equilibrium statistical mechanics, 
the system described by the Hamiltonian (\ref{hfi4d}) has a 
critical point at finite temperature provided that $d>1$. 
In the 
following we will restrict ourselves to the case $d=2$; the case
$d=3$ has been considered elsewhere, together with the cases
of $O(n)$-invariant $\varphi^4$ models \cite{Gatto}.

\subsection{Dynamics and thermodynamic observables}

Our dynamical approach to the model defined by the 
Hamiltonian (\ref{hfi4d}) is based on the direct solution of 
the equations of motion --- Hamilton's equations --- that 
read as
\bea
\dot{\varphi}_i & = & \pi_i, \nonumber \\
\dot{\pi}_i & = & 
J\sum_{\mu=1}^{d}(\varphi_{i+\mu}+\varphi_{i-\mu} - 2 
\varphi_{i}) + 
m^2 \varphi_i - \frac{1}{3!}\lambda \varphi_i^3~. 
\label{eqmoto}
\eea
The numerical integration of the $2N$ equations
(\ref{eqmoto}) has been performed by means of a third-order bilateral 
symplectic algorithm \cite{algo}.
The parameters have been chosen as follows: 
$J=1$, $m^2=2$, $\lambda=0.6$.
\label{sceltaparametri} 
The average of any observable is defined as a time average, 
i.e. 
\beq
\langle f \rangle = \lim_{t \to \infty}\frac{1}{t} \int_0^t 
f(\varphi(\tau),\pi(\tau))\, d\tau\,\,;
\eeq
In practice, such an average is evaluated by means of a
discrete sampling of $f$.

As already discussed in Sec.\ \ref{sec_general}, even if we 
cannot rigorously prove that the invariant ergodic measure 
associated with the Hamiltonian dynamics of our system is 
the microcanonical measure, nevertheless the microcanonical 
ensemble is the statistical ensemble which is naturally 
associated with Hamiltonian dynamics, for it is defined 
directly from the dynamics itself. Thus, in defining the 
dynamical observables which represent the thermodynamic 
properties of our model, we shall consider the 
microcanonical ensemble. 

The phase space density of the microcanonical measure 
can be written as \cite{Khinchin} 
\beq
\varrho_{\rm micro} = \frac{1}{\omega} \delta\left({\cal 
H}(\varphi,\pi) - E\right)
\eeq
where the normalization $\omega$ is given by 
\beq
\omega = \int \delta\left({\cal H}(\varphi,\pi) - E\right)\,  
d\varphi\, d\pi\,\, ,
\label{omega}
\eeq
and $d\varphi\, d\pi$ is a shorthand for $d\varphi_1\cdots 
d\varphi_N d\pi_1 \cdots d\pi_N$.

The entropy $S$ is defined once the 
external parameter $E$ (the energy of the system) is assigned. 
Different definitions of $S$ can be given, all of which are 
equivalent in the thermodynamic limit. The two common
definitions are (we set $k_B = 1$) 
\bea
S^{\Omega} (E) & = & \log \Omega(E) \,\, ,\label{entropy}\\
S^{\omega} (E) & = & \log \omega(E) \,\, ,
\eea
where 
\beq
\Omega (E) = \int \vartheta\left({\cal H}(\varphi,\pi) - 
E\right)\,  d\varphi\, d\pi\,\, .
\label{Omega}
\eeq
The temperature is defined thermodynamically as 
\beq
\beta = \frac{1}{T} = \frac{\partial S(E)}{\partial E} 
\,\, ;
\label{T_def}
\eeq
in the above definition, $S$ can denote either $S^\Omega$ or 
$S^\omega$, whence we have two different definitions of 
temperature, $T^\Omega$ and $T^\omega$. Again the two 
definitions are equivalent in the thermodynamic limit. Let 
us now limit ourselves to the case of natural dynamical 
systems, i.e., to Hamiltonian systems with $N$ degrees of 
freedom whose Hamiltonian can be written as 
\beq
{\cal H} = \sum_{i=1}^N \frac{\pi_i^2}{2m} + V(\varphi) = K 
+ V\,\, ;
\eeq
then the temperature can be expressed in terms of 
the kinetic energy  as follows \cite{Pearson}:
\bea
T^\Omega & = & \frac{\Omega}{\omega} \,\, = \,\, \frac{2}{N} 
\langle K \rangle\,\, ; 
\label{T_Omega} \\
T^\omega & = & \left[ \left( \frac{2}{N} - 1 \right) \langle 
K^{-1} \rangle \right]^{-1} \,\, .
\label{T_omega} 
\eea
It is worth noticing that for natural dynamical systems the 
definition (\ref{T_Omega}) also coincides with the 
definition of temperature in the canonical ensemble. In what 
follows, unless explicitly stated otherwise, we shall adopt 
this definition of temperature and we shall drop the superscript 
$\Omega$, i.e., $T = T^\Omega$. 
A more general approach to microcanonical thermodynamics
\cite{Rugh} has been recently introduced, 
which allows to find new expressions for
thermodynamical observables. Anyhow, the definition (\ref{T_Omega}) 
is the best suited for numerical simulations in the case of natural systems
\cite{Giardina}.

Another thermodynamic quantity to be computed in our 
simulations is the constant-volume specific heat, 
defined as
\beq
\frac{1}{C_v} = \frac{\partial T(E)}{\partial E} \,\, .
\eeq
From the definition and Eq. (\ref{T_Omega}) it follows that 
\cite{Pearson}
\beq
{c_v} = \frac{C_v}{N} = \left[ N - (N - 2) \langle K \rangle 
\langle K^{-1} \rangle \right]^{-1} \,\, ;
\label{cv_micro}
\eeq
in the thermodynamic limit $N \to \infty$ this expression 
for the specific heat reduces to the well-known 
Lebowitz-Percus-Verlet formula \cite{LPV},
\beq
c_v = \frac{1}{2}\left( 1 - \frac{N}{2} 
\frac{\langle K^2 \rangle - \langle K \rangle^2}{\langle K 
\rangle^2}\right)^{-1}~,
\label{cv_lpv}
\eeq
which is almost universally used to compute the specific 
heat in molecular-dynamics simulations \cite{Frenkel}.
Notice that Eq. (\ref{cv_lpv}) is derived 
from an asymptotic expansion of the microcanonical fluctuations 
in terms of the canonical ones, thus it is valid only in the 
limit of large $N$. On the contrary, the formula 
(\ref{cv_micro}) is exact at {\em any} value of $N$. Hence 
this is the correct expression to be used in a finite 
system. 

Finally, we will consider the order parameter, i.e. the 
``magnetization'' 
\beq
\langle \varphi \rangle = \frac{1}{N} \left\langle 
\sum_{i=1}^N \varphi_i \right\rangle \,\, .
\eeq

\section{Results of the dynamical simulations}

\label{sec_results}

We now turn to the problem of detecting the phase transition 
in our lattice $\varphi^4$ Hamiltonian system. According to 
the thermodynamic definition, we must look for a singularity 
of the thermodynamic observables as functions of the energy 
--- or better as a function of the energy density 
$\varepsilon = E/N$ which remains finite as $N \to \infty$ and 
facilitates the comparison of the results obtained 
at different lattice
sizes. In particular, we look for a singularity in 
$c_v(\varepsilon)$. Besides, being the transition 
associated with the spontaneous breaking of the ${\bf Z}_2$ 
symmetry, we will look for the appearance of a nonzero order 
parameter, i.e. for a nonvanishing value of the average 
magnetization $\langle \varphi \rangle$.

\subsection{Binder cumulants}

In the canonical ensemble, a phase transition may occur 
only in the thermodynamic limit. As long as $N$ is finite, 
all the thermodynamic quantities are regular functions of 
the temperature, and ergodicity and symmetry are not broken.
Nevertheless, some marks of the transition clearly show up 
also in a finite system. The specific heat does not diverge, 
but exhibits a peak --- whose height grows with the size of 
the system --- at a temperature $T_c^{c_v}(N)$, while
the order parameter is expected to be vanishing in the whole 
temperature range at any finite value of $N$. Neverthless 
this is true only in principle: in practice, e.g. in a 
canonical MonteCarlo simulation, where the simulation time 
is necessarily finite, the system is 
trapped in one of the two phases for a ``time'' which grows 
exponentially with $N$ \cite{Goldenfeld}, and one observes a 
fictitious symmetry breaking at a temperature 
$T_c^{\varphi}(N)$. The latter temperature in general does 
not coincide with $T_c^{c_v}(N)$, even if
\beq
\lim_{N \to \infty} T_c^{c_v}(N) = \lim_{N \to \infty} 
T_c^{\varphi}(N) = T_c^\infty \,\, .
\eeq

In the microcanonical ensemble ergodicity breaking may occur 
also at finite $N$, hence one could expect a ``true'' 
critical energy to be defined also at finite $N$. No 
rigorous theoretical result is at disposal as far as 
this aspect is concerned. Nevertheless, it is reasonable to expect --- 
and this is indeed what is observed  --- 
that the actual behaviour of the thermodynamic 
functions will be similar to the canonical case, at least as 
$N$ is sufficiently large. In particular, we expect the 
specific heat to exhibit a peak at a critical energy density 
which is a function of $N$, and the order parameter to be 
nonvanishing below another critical energy, again depending 
on the size of the system.

In the framework of the statistical theory of critical 
phenomena, finite-size scaling \cite{Dunweg} allows to 
estimate the critical properties of the infinite system from 
the values of the thermodyamic observables in finite samples 
of different sizes. In particular it is possible to locate 
the critical point by means of the so-called {\em Binder 
cumulants} \cite{Binder}. The Binder cumulant $g$ is defined 
for our system as
\beq
g = 1 - \frac{\langle \varphi^4 \rangle}{3\langle \varphi^2 
\rangle^2}\,\, .
\label{binder}
\eeq
In the disordered phase the probability distribution of the 
order parameter will be nearly Gaussian with zero mean, 
hence $g \simeq 0$. At variance, at zero temperature (or 
energy), when $\varphi_i \equiv \varphi_0$ with no 
fluctuations, $g = 2/3$. At different sizes of the system, 
$g(T)$ will decay from $2/3$ to $0$ 
with different patterns. It is remarkable that the 
value of $g$ at $T_c^{\infty}$ is {\em independent} of $N$, 
provided $N$ is large enough for the scaling regime to set 
in, hence the critical point is given by 
the intersection of the different curves $g(T)$ 
for different values of $N$. In principle, two different 
sizes are sufficient to locate a transition; in practice, 
owing to the unavoidable numerical errors which affect $g$, 
it will be necessary to consider three or more values of 
$N$. Moreover, the value of $g$ at the critical point, 
usually referred to as $g^*$, is a universal quantity; 
for a simple proof see e.g. Ref. \cite{Dunweg}. 

The theory behind the Binder cumulant method is totally 
internal to canonical statistical mechanics: to our 
knowledge, no extension of this theory to the microcanonical 
ensemble exists. Nevertheless we will adopt the pragmatic 
point of view of assuming its validity as a numerical tool 
also in our dynamical simulations, and our operative 
definition of the critical energy density 
$\varepsilon_c^\infty$ will be the intersection point of the 
curves $g(\varepsilon)$ at different $N$. The consistency of 
the method will be checked {\em a posteriori}. In the 
following, unless explicitly stated otherwise, 
$\varepsilon_c$ and $T_c$ will denote respectively 
$\varepsilon_c^\infty$ and $T_c^\infty$.

\begin{figure}
\centerline{\psfig{file=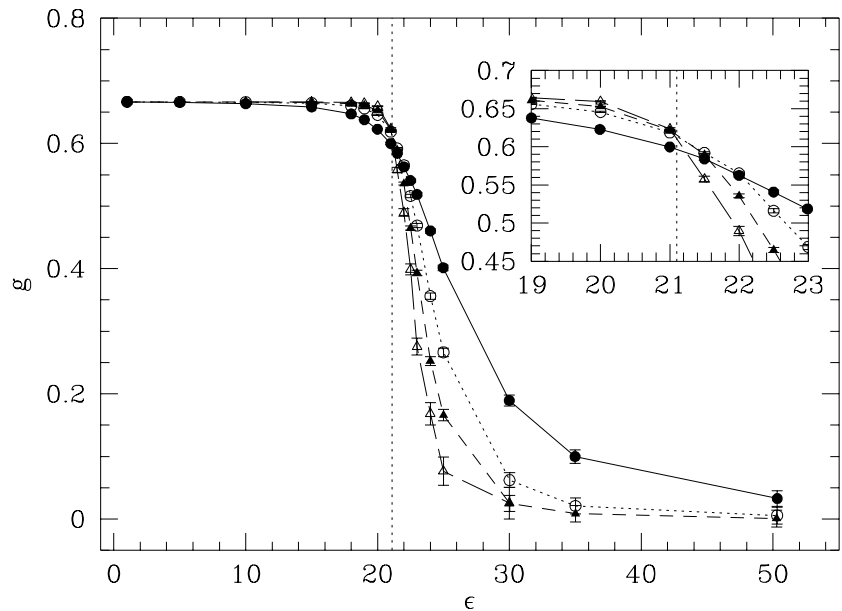,height=9cm,clip=true}}
\caption{Binder 
cumulants $g$ (\protect\ref{binder}) vs. energy density 
$\varepsilon$ for different sizes of the system. The symbols 
denote respectively $N = 10^2$ (solid circles), $N = 20^2$ 
(circles), $N = 30^2$ (solid triangles), $N = 50^2$ (triangles).
The vertical dotted line marks the estimated value of 
$\varepsilon_c \simeq 21.1$. The inset shows a magnification of the 
transition region.}
\label{fig_binder}
\end{figure}

The results for $g(\varepsilon)$ at different sizes for the 
two-dimensional lattice $\varphi^4$ model are shown in Fig. 
\ref{fig_binder}. The crossing of the various 
curves at $\varepsilon_c \simeq 21.1$ is evident. 

\subsection{Thermodynamical observables}

\subsubsection{Temperature}

\begin{figure}
\centerline{\psfig{file=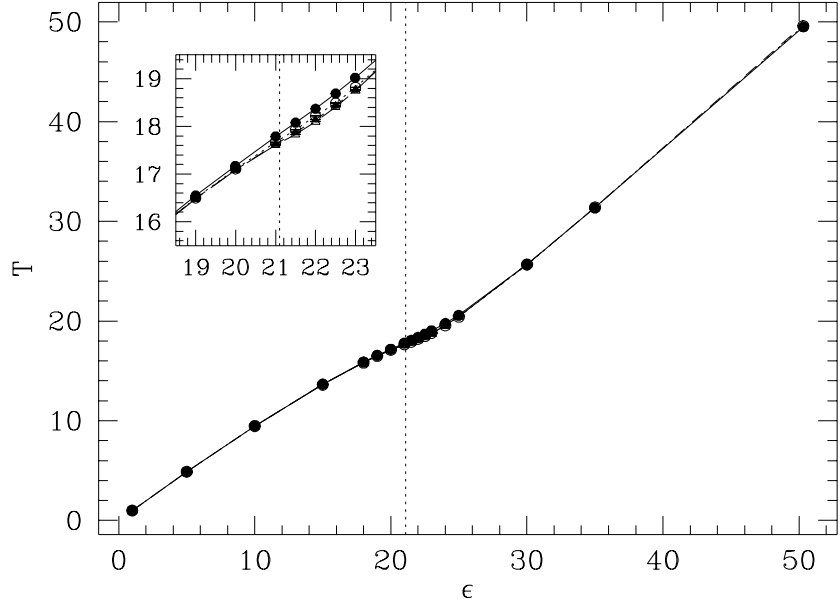,height=9cm,clip=true}}
\caption{Temperature $T$ vs. energy density 
$\varepsilon$. Symbols as in Fig. \protect\ref{fig_binder}.}
\label{fig_temp}
\end{figure}

The temperature of the two-dimensional $\varphi^4$ system, 
numerically determined according to Eq. (\ref{T_Omega}), is 
plotted in Fig. \ref{fig_temp}. 
Notice the change in the convexity of the function $T(\varepsilon)$ at
$\varepsilon = \varepsilon_c$.

\begin{figure}
\centerline{\psfig{file=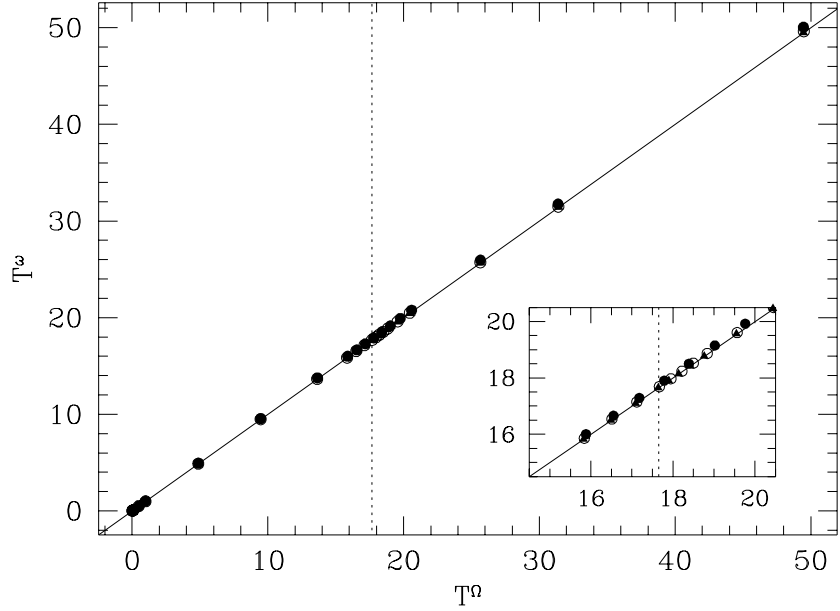,height=9cm,clip=true}}
\caption{Comparison between the two different definitions of 
temperature $T = T^\Omega$ --- Eq. (\protect\ref{T_Omega}), 
solid circles --- and $T = T^\omega$ --- Eq. 
(\protect\ref{T_omega}), squares --- at three different 
lattice sizes: $N=10^2$ (solid circles), $N=20^2$ (circles), 
$N=30^2$ (solid triangles). The solid line is the line 
$T^\omega = T^\Omega$, the dotted vertical line marks the 
estimated value of $T_c$ (see Fig. \protect\ref{fig_binder_T}).}
\label{fig_temp_confr}
\end{figure}

In Fig. \ref{fig_temp_confr} a comparison between the 
temperatures computed according to the two definitions 
(\ref{T_Omega}) and (\ref{T_omega}) is shown. It is evident 
that already in a $10 \times 10$ lattice the two 
temperatures are practically identical. 

\begin{figure}
\centerline{\psfig{file=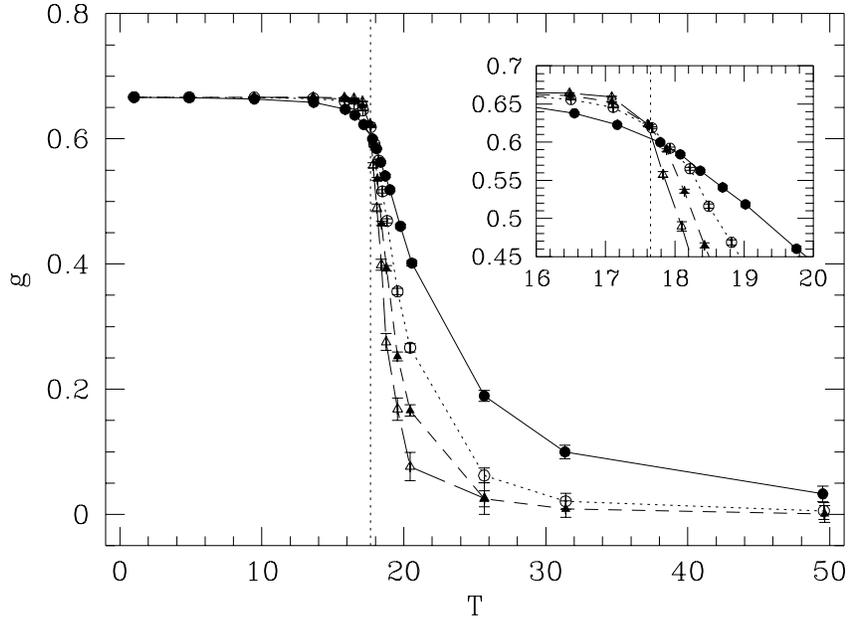,height=9cm,clip=true}}
\caption{Binder 
cumulants $g$ (\protect\ref{binder}) vs. temperature $T$. 
Symbols as in Fig. \protect\ref{fig_binder}. The vertical 
dotted line marks the estimated value of 
$T_c \simeq 17.65$. The inset shows a magnification of the 
transition region.}
\label{fig_binder_T}
\end{figure}

In Fig. \ref{fig_binder_T} the Binder cumulants are plotted 
vs. temperature $T$ to locate the critical temperature 
$T_c$. We see that this $T_c$ is consistent with $T(\varepsilon_c)$.

\subsubsection{Specific heat}

\begin{figure}
\centerline{\psfig{file=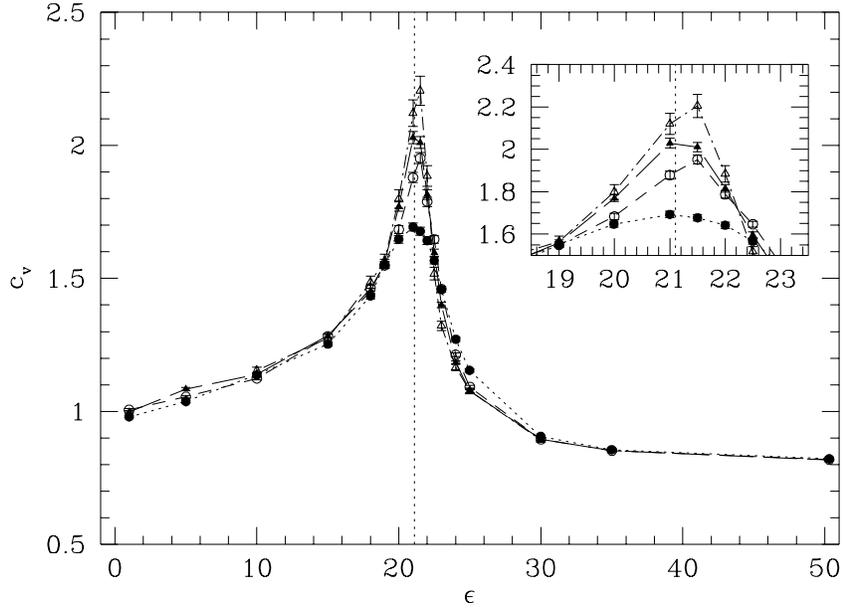,height=9cm,clip=true}}
\caption{Specific 
heat $c_v$ vs. energy density 
$\varepsilon$. Symbols as in Fig. \protect\ref{fig_binder}.}
\label{fig_cv}
\end{figure}

The specific heat of the two-dimensional $\varphi^4$ system 
is plotted against energy density in Fig. \ref{fig_cv}.
The asymptotic values of the specific heat in the limits 
$\varepsilon \to 0$ and
$\varepsilon \to\infty$ are exactly known. In fact at low 
energies the anharmonic terms in the Hamiltonian can be 
neglected, the system behaves as a collection af harmonic 
oscillators and $c_v \approx 1$. In the high-energy limit 
the quadratic terms in the potential are negligible with 
respect to the quartic ones, whence  
$c_v \approx 1/2+1/4=3/4$. At intermediate 
energy densities, a neat peak shows up whose position is close 
to $\varepsilon_c$. The height of the 
peak grows with $N$. 

\subsubsection{Order parameter}

\begin{figure}
\centerline{\psfig{file=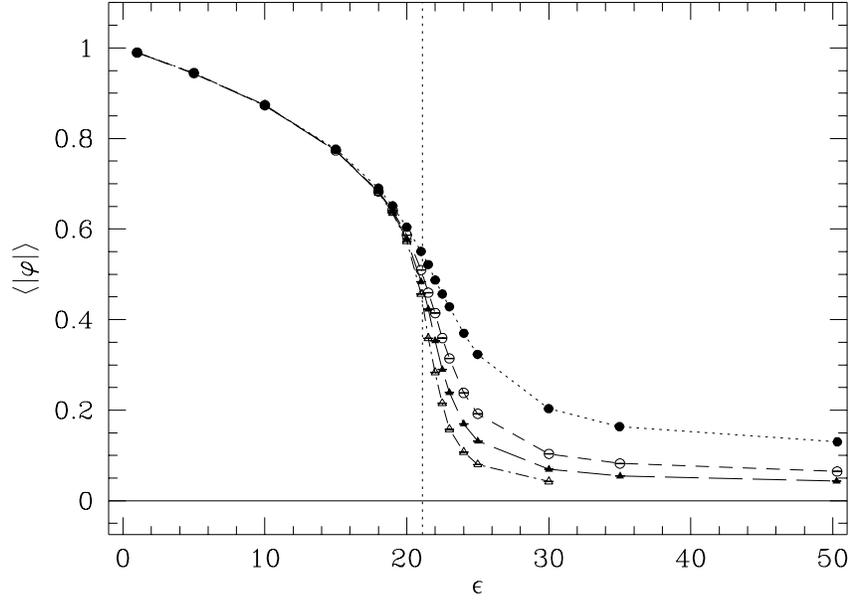,height=9cm,clip=true}}
\caption{Absolute magnetization
$\langle |\varphi| \rangle$ vs. energy density 
$\varepsilon$. Symbols as in Fig. \protect\ref{fig_binder}.}
\label{fig_abs_phi}
\end{figure}

Let us turn to the behaviour of the order parameter $\langle \varphi \rangle$. 
In principle, a nonzero value of $\langle \varphi \rangle$ is the
characteristic
signal of the breaking of the ${\bf Z}_2$ symmetry, which is in turn a
consequence of ergodicity breaking. In practice, as long as a finite
system is considered, the situation is more subtle. We have already observed
that in a canonical ensemble,
where the temperature is fixed, ergodicity can be broken only in the
thermodynamic limit, hence the eventual appearence of a nonzero order
parameter is a consequence of the necessarily finite observation time.
In order to get reliable results the standard procedure is then to compute
$\langle |\varphi| \rangle$ rather than $\langle \varphi \rangle$; in this
way one has a quantity whose average is always nonnegative, and
that coincides with the ``true'' order parameter as $N \to \infty$.
Obviously in the symmetric phase this quantity, at finite $N$, is nonzero:
its amplitude will decrease as $1/\sqrt{N}$. This is precisely 
the behaviour of our numerical results reported in Fig. \ref{fig_abs_phi}.

\begin{figure}
\centerline{\psfig{file=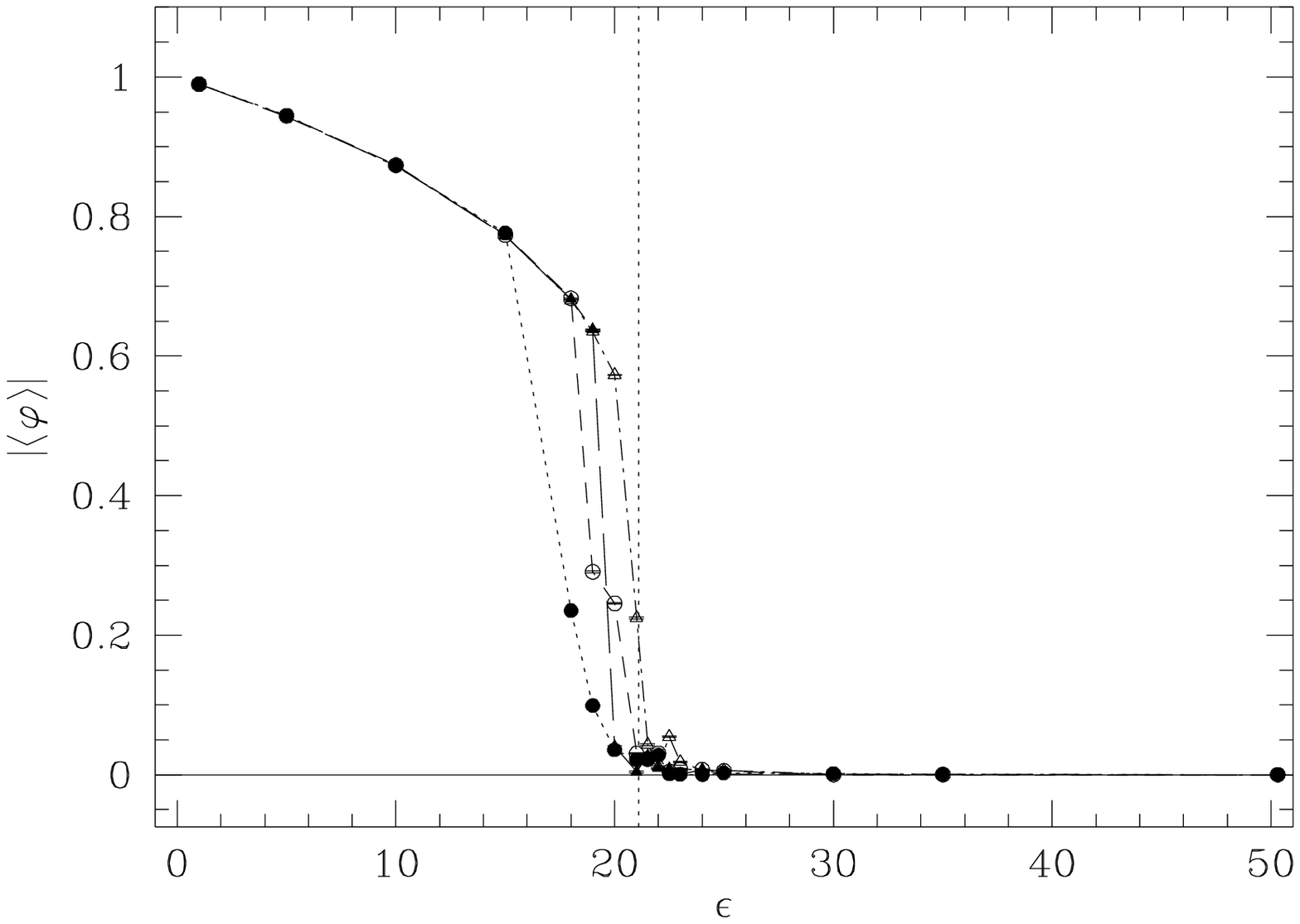,height=9cm,clip=true}}
\caption{Magnetization
$\langle\varphi\rangle$ vs. energy density 
$\varepsilon$. Symbols as in Fig. \protect\ref{fig_binder}.}
\label{fig_phi}
\end{figure}

In the microcanonical ensemble
the situation is more complicated, in fact ergodicity breaking is
no longer forbidden at finte $N$, hence a nonzero value of 
$\langle \varphi \rangle$ might be either a finite-time artifact or
a ``true'' signal of symmetry breaking. Anyhow, the mere observation of the
result is not sufficient to discriminate between these two
alternatives.  The order parameter (the ``true'' one, whose absolute
value is taken only {\em after} the average for graphical reasons) is
reported in Fig. \ref{fig_phi}. We see that at each $N$ there
is a value of $\varepsilon$ below which the symmetry appears to be broken,
and the value of $\langle \varphi \rangle$ seems to move almost abruptly
from zero to a finite value. These ``critical'' energy densities 
are the closer to $\varepsilon_c$ the larger $N$ is.

\subsection{Critical behaviour}

The classical lattice $\varphi^4$ model, whose Hamiltonian 
is invariant under a discrete ${\bf Z}_2$ symmetry, belongs 
to the universality class of the Ising model. 

\begin{figure}
\centerline{\psfig{file=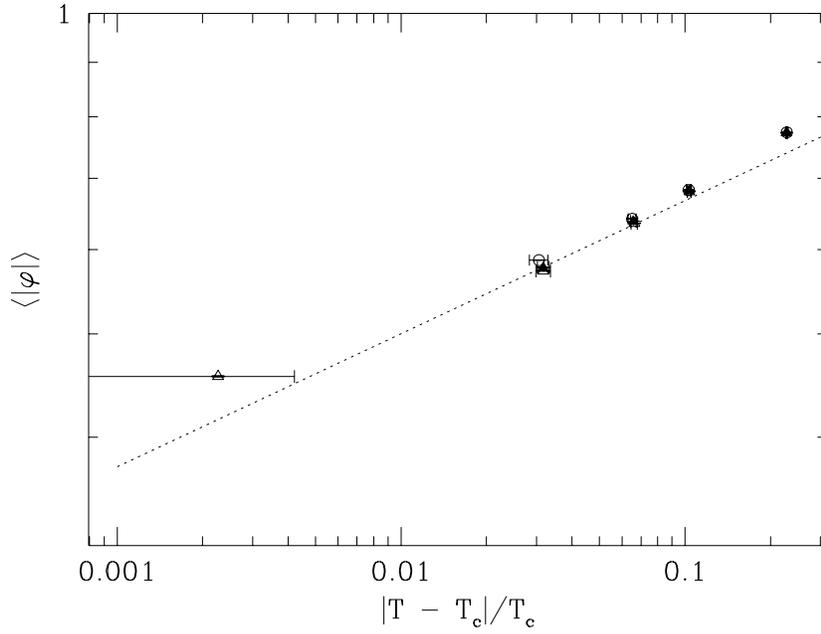,height=9cm,clip=true}}
\caption{Scaling behaviour of the order parameter  
in the two-dimensional $\varphi^4$ model. Symbols as in 
Fig. \protect\ref{fig_binder}; only $20^2$, $30^2$ and $50^2$ lattices
are considered. The dotted line is the exact 
result for the Ising model, i.e., the power law $|T-T_c|^{1/8}$.}
\label{fig_scaling_phi}
\end{figure}

\begin{figure}
\centerline{\psfig{file=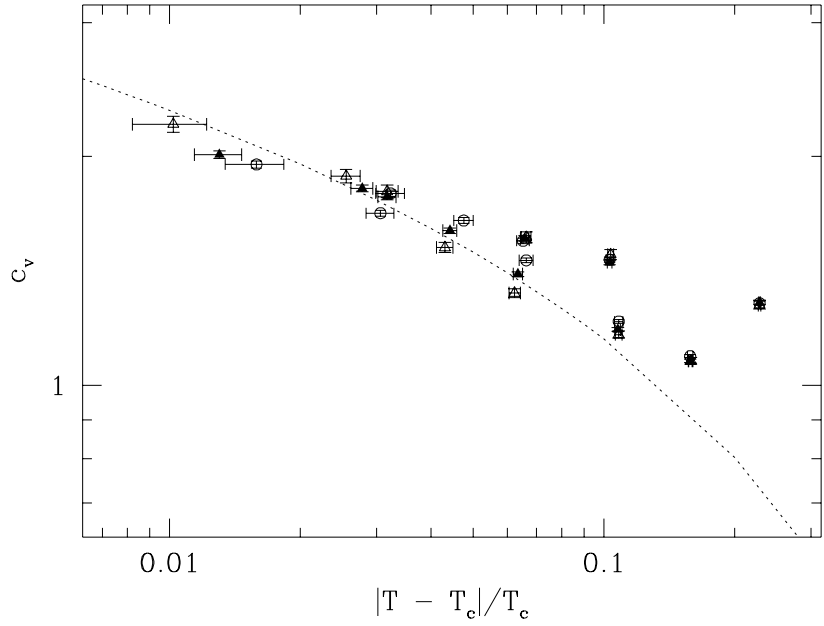,height=9cm,clip=true}}
\caption{The same as in Fig. \protect\ref{fig_scaling_phi} 
for the specific heat. Here the dotted line is the logarithmic behaviour.}
\label{fig_scaling_cv}
\end{figure}

Our goal is not to obtain a precise measurement of the  
critical exponents, but only to check the consistency of our 
dynamical results with the statistical theory. The fact that 
the $\varphi^4$ theory belongs to the Ising universality 
class is a great advantage, because the critical exponents 
of the Ising model in two dimensions have been computed  
exactly: in particular, the order parameter critical
exponent is $\beta = 1/8$ and the specific heat one is $\alpha = 0$
(the specific heat has a logarithmic singularity).
We can compare the numerical outcomes of our simulations 
with the predicted Ising values in order to check whether 
the dynamically simulated critical behaviour is compatible 
with the predictions of statistical mechanics. In Figs. 
\ref{fig_scaling_phi} and \ref{fig_scaling_cv} the scaling 
behaviours of the order parameter and of the specific heat are
compared to the exact Ising behaviours in two dimensions. 
The results are clearly very well compatible with the theory.
This is a sign that the dynamical approach effectively
reproduces the thermodynamical phase transition of the 
$\varphi^4$ model.
 
\section{Dynamical properties}

Up to now we have shown that the outcomes of the dynamical numerical
simulations are perfectly consistent, both qualitatively and quantitatively, 
with the theoretical expectations regarding the phase transition of
the $\varphi^4$ model.  
Though obtained through dynamics, all these results deal with 
equilibrium time averages: the time variable,
even if not eliminated from the very beginning as in the statistical
approach, has been integrated out in the averaging procedure. 
Now, we can also wonder whether there are intrinsically dynamical properties
of our system that are relevant for the phase
transition itself. Moreover, since from our point of view 
ergodicity breaking has its origin in the dynamics, we can try to 
understand what features are associated to a Hamiltonian ergodicity
breaking.
 
\subsection{Time correlation functions}

\begin{figure}
\centerline{\psfig{file=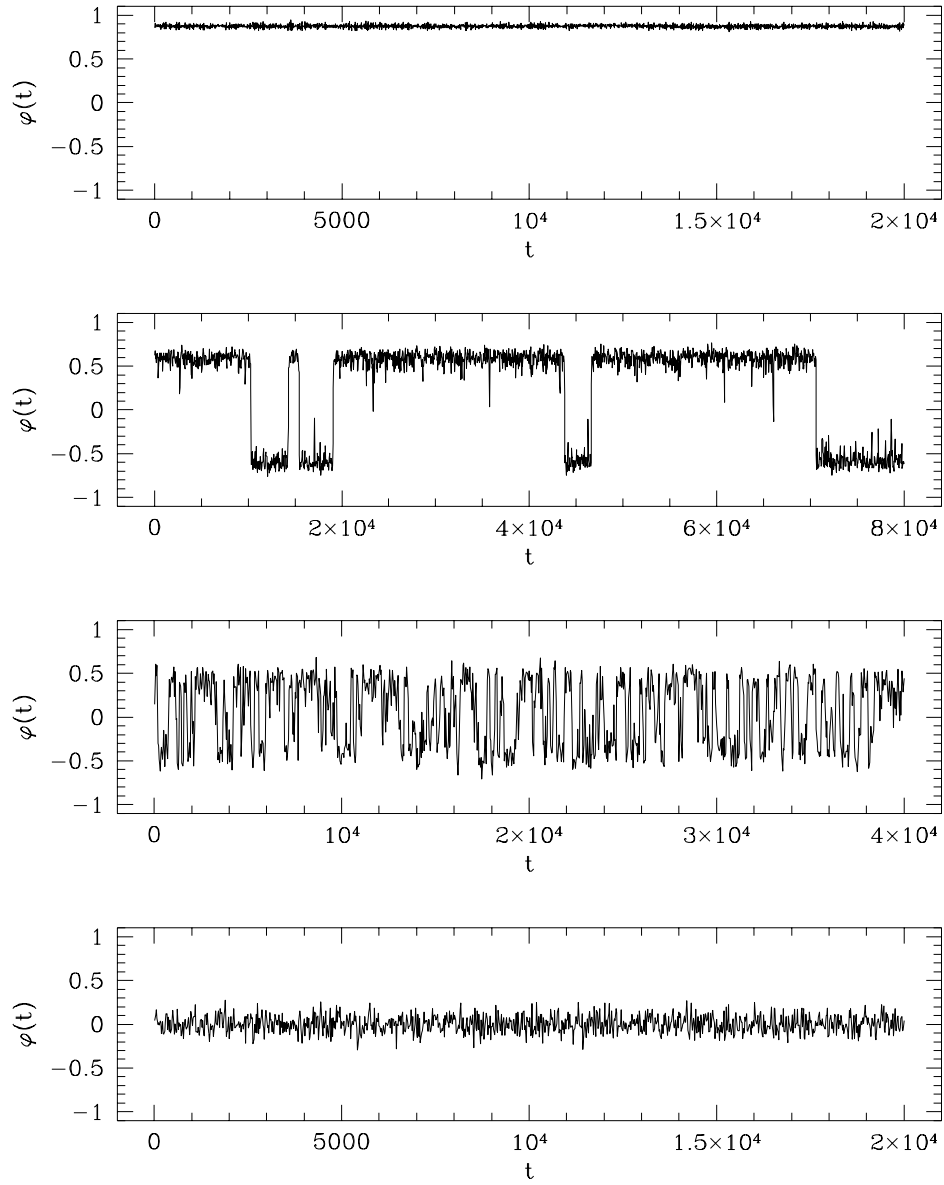,height=18cm,clip=true}}
\caption{Temporal behaviour of the order parameter $\varphi(t)$
for a $20^2$ lattice at different values of $\varepsilon$: from
top to bottom, $\varepsilon = 10$, 20, 22.5 and 35.}
\label{fig_phi_time}
\end{figure}

\begin{figure}
\centerline{\psfig{file=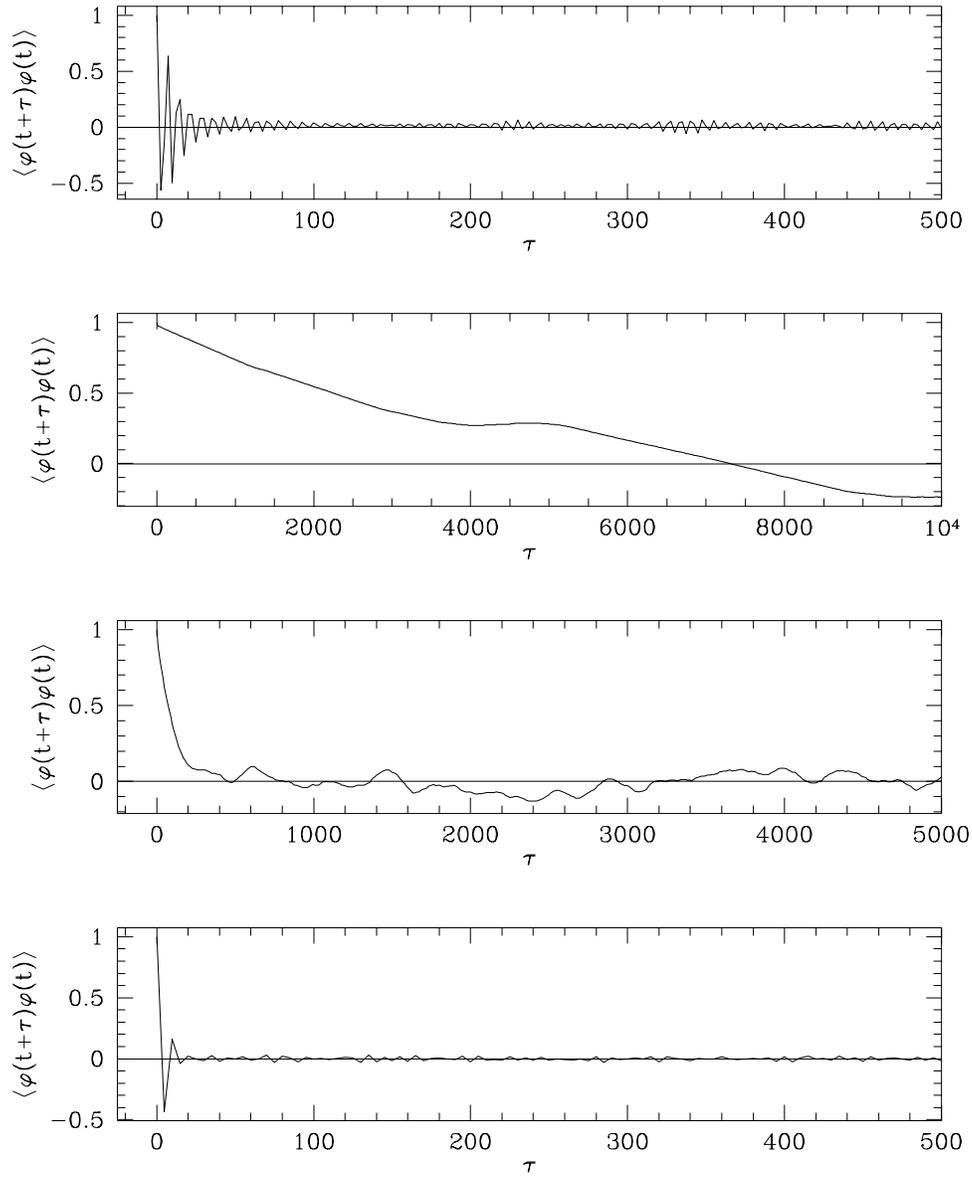,height=18cm,clip=true}}
\caption{Time autocorrelation function of the order parameter
$\varphi(t)$ for the same lattice and at the same
energies as in Fig. \protect\ref{fig_phi_time}.}
\label{fig_acf}
\end{figure}

The first dynamical property that we are going to study is the dynamics
of the order parameter $\varphi(t)$. A qualitative understanding of what is 
going on is already provided by the time series $\varphi(t)$ itself:
some examples are reported in Fig. \ref{fig_phi_time}, for various values
of $\varepsilon$, in the case of a $20 \times 20$ lattice.
But a much more interesting information is contained in the time 
correlation functions of the order parameter,
\beq
C_\varphi(\tau) = \frac{\langle \varphi(t) \varphi (t+\tau) \rangle}
{\langle \varphi^2(t) \rangle}~;
\eeq
some of these functions are plotted in Fig. \ref{fig_acf}. We  
immediately note that close to the critical energy --- and indeed very close to
the ``finite $N$'' critical energy where ergodicity appears to be
broken according to Fig. \ref{fig_phi} --- the shape of the correlation
function changes rather sharply from an oscillatory pattern with a 
superimposed decay to a pattern 
indicating that the values of $\varphi(t)$ are correlated
over an extremely long period of time.
Such a phenomenon is obviously reminiscent of {\em critical slowing
down}. The latter is a feature of the local\footnote{There are
indeed ``smart'' dynamics which greatly reduce, or completely eliminate,
the critical slowing down; the common feature of these dynamical rules is
that of being highly nonlocal \protect\cite{Sokal}.}
phenomenological dynamical evolutions, e.g., Monte Carlo dynamics, 
constructed in order
to have the Boltzmann distribution $e^{-\beta {\cal H}}$ as limiting
invariant probability distribution, when the system described by ${\cal H}$
is close to a continuous phase transition. The usual explanation for
the critical slowing down is that it is a consequence of the divergence
of the static correlation length at the phase transition. However
we are now considering a deterministic Hamiltonian microscopic dynamics
and not a phenomenological MonteCarlo dynamics, 
hence our results
clearly show that some kind of critical slowing down exists
also in the microscopic natural dynamics. Something similar, i.e., the
development of low-frequency collective oscillations, was observed
in a planar mean-field Heisenberg model close to criticality \cite{Ruffo}.
This result suggests that 
a Hamiltonian description of critical slowing down is possible and it might be
very useful in dynamically understanding the phenomenon of ergodicity
breaking (the study of a simple model which provides 
a first step towards this Hamiltonian approach to critical slowing
down is presented in Ref. \cite{thesis}). 

\begin{figure}
\centerline{\psfig{file=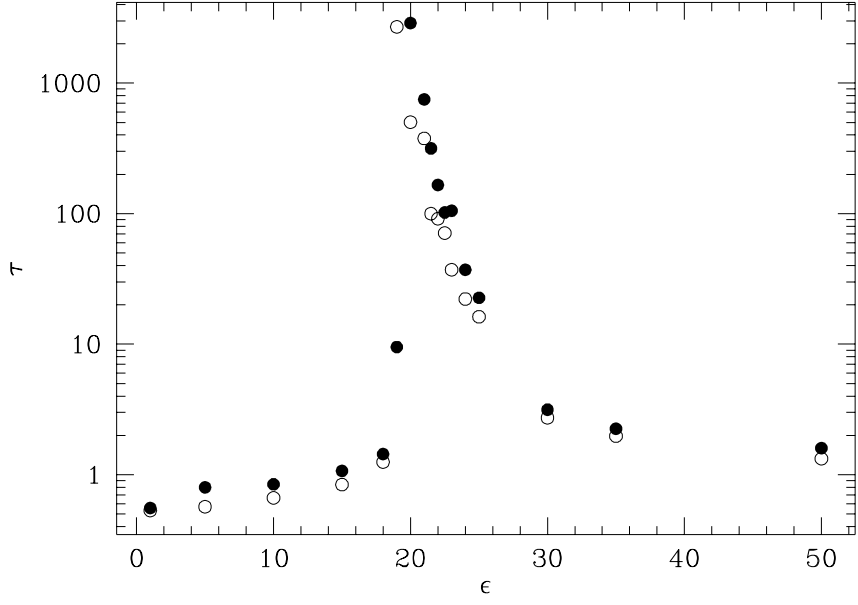,height=9cm,clip=true}}
\caption{Characteristic time $\tau$ of the order parameter $\varphi(t)$
(see text). Circles: $10^2$ lattice, solid circles: $20^2$ lattice.}
\label{fig_tau}
\end{figure}

In order to obtain a synthetic information from the time correlation functions,
let us define a characteristic time $\tau$ as follows:
\beq
\tau = \int_0^{t_0} C_\varphi(t) \,dt~,
\label{tau_correlation}
\eeq
where $t_0$ is the time where $C_\varphi$ has its first zero. Such a definition
is to a large extent arbitrary, nevertheless it provides a
relevant time scale either if the correlation function is
oscillatory with typical frequency $\omega$ and with only a weak damping 
--- as it happens at low energy --- in which case $\tau \approx \omega^{-1}$,
or in the case of an exponentially decaying correlation with inverse 
time constant
$\gamma$, in which case $\tau \approx \gamma^{-1}$. The values of $\tau$ 
computed with Hamiltonian dynamics are reported in Fig. \ref{fig_tau}.
The striking result is that the characteristic time is rapidly growing
(notice the logarithmic vertical axis) as the system approaches the phase
transition: the position of the peak is close to the ``finite $N$''
critical energy.

\subsection{Chaotic dynamics}

The lattice $\varphi^4$ model is a nonintegrable dynamical 
system. In the two limits $\varepsilon \to 0$ and 
$\varepsilon \to \infty$, the system is integrable. The 
two integrable limits respectively represent a system of coupled 
harmonic oscillators and a system of independent quartic 
oscillators. The dynamics will be always chaotic in the 
whole energy range. Nevertheless, in analogy with other 
nonlinear oscillator systems, we expect that as the energy 
density is varied there exist different dynamical regimes 
characterized by different behaviours of the Lyapunov 
exponent $\lambda$. In particular, the following questions 
naturally arise. Is there any peculiar behaviour of the 
Lyapunov exponent in correspondence with the phase 
transition? Is there any transition between different
chaotic regimes in the $\varphi^4$ lattice model, 
and, if yes, is there any relationship between these 
different dynamical regimes and the thermodynamic phases?

There are not yet general and
conclusive answers to these questions. Even if the study of 
a possible relation between chaos and phase transitions is a 
rather recent issue, which started with the already mentioned 
pioneering work by Butera and Caravati \cite{ButeraCaravati},
very different results have already 
appeared in the literature, ranging from the claim of the 
discovery of a ``universal'' divergence in $\lambda$ as the 
system approaches criticality in a class of models 
describing clusters of particles \cite{Rapisarda}, to the 
observation that the Lyapunov exponent attains its minimum 
in correspondence with the phase transition in Ising-like 
coupled map lattices \cite{Duke_pre}, to the apparent 
insensitivity to the liquid-solid phase transition of the  
Lyapunov spectra of hard-sphere and Lennard-Jones systems 
\cite{Dellago}; for other recent results on
first-order transitions see Ref. \cite{Mehra}.

More recently, some very interesting results have been obtained
concerning mean-field-like models, in particular globally coupled
rotators. Numerical results \cite{RuffoRapisarda},
though of not easy interpretation due to 
strong finite-size effects, indicate that in these systems,
that undergo a mean-field phase transition at a critical energy 
density $\varepsilon_c$, the Lyapunov exponent vanishes in the whole
disordered phase, whereas it is positive in the low-energy (ordered) phase. 
Such a result has been theoretically confirmed in 
a very recent work \cite{Firpo}. 
Since the latter work is based on the 
application of the theoretical tools described in Sec.
\ref{sec_geometry}, 
we will discuss there its results.

\begin{figure}
\centerline{\psfig{file=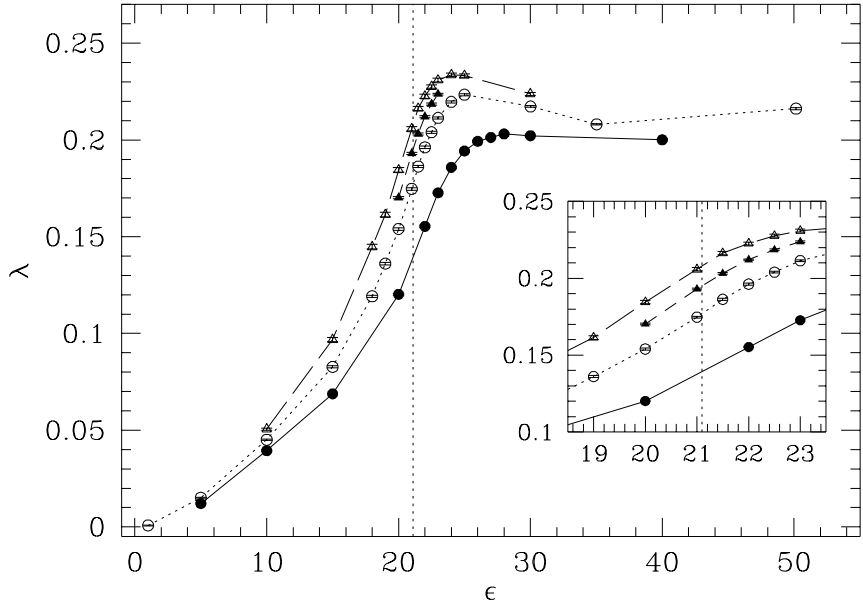,height=9cm,clip=true}}
\caption{Lyapunov exponent $\lambda$ vs. energy density 
$\varepsilon$. Symbols as in Fig. \protect\ref{fig_binder}.}
\label{fig_lyap}
\end{figure}

\begin{figure}
\centerline{\psfig{file=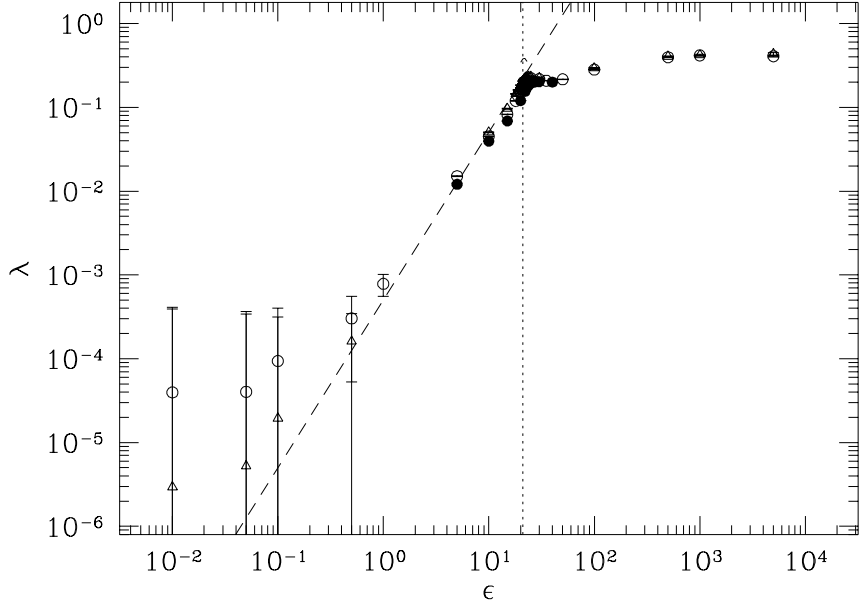,height=9cm,clip=true}}
\caption{The same as in Fig. \protect\ref{fig_lyap} in
a wider energy range and using logarithmic scales. The dashed line is
the power law $\lambda \propto \varepsilon^2$.}
\label{fig_lyap_log}
\end{figure}

Our simulation results are plotted in Fig. \ref{fig_lyap}.
The first numerical evidence is that there is a strong
dependence of $\lambda$ on $N$, what is peculiar of the
presence of a phase transition. Moreover at large $N$ a maximum
of $\lambda(\varepsilon)$ develops which eventually seems 
to move towards the critical energy density. Nevertheless
no sharp, or singular, transition between different behaviours 
is found near $\varepsilon_c$, at variance with the
three-dimensional case \cite{Gatto}; it is possible that such 
a sharp transition shows up 
as $N \to \infty$, but no clear indications of this fact are provided 
by our results. Nevertheless the behaviour of $\lambda(\varepsilon)$ 
in the region $\varepsilon < \varepsilon_c$ is
very different from that of the thermodynamically disordered
region, i.e, in the former $\lambda$ rapidly grows with $\varepsilon$,
while in the latter it has a quasi-flat shape (which is expected
to become a decrease as $\varepsilon$ is large enough, because the 
high-energy limit is a integrable limit for the model).
This behaviour is more clearly seen in Fig. \ref{fig_lyap_log} 
where a wider energy range is considered and logarithmic axes
are used.
This suggests that 
the phase transition has a dynamical counterpart in a 
passage between different chaotic regimes. 
However --- at present --- this statement cannot be formulated in 
a conclusive way, mainly because there is no 
clear and unique way to define and characterize a transition 
between different chaotic regimes (a recent
improvement towards an unambiguous characterization of this
kind of transitions can be found in Ref. \protect\cite{Giardina}).

Moreover, numerical experiments show that 
the detailed behaviour of the Lyapunov exponent 
close to the transition
does not show ``universal'' features, i.e., it depends on the 
details of the Hamiltonian: on the contrary, 
in the low-energy range also for the $\varphi^4$ model
we have $\lambda \propto \varepsilon^2$ that is the same behaviour
observed in most, if not all, of the systems of coupled oscillators. 
Hence it is still unclear which feature  
of the $\varepsilon$-dependence of the Lyapunov 
exponent has to be related with the phase transition.
An exception is the already mentioned case of the mean-field rotators
model \cite{RuffoRapisarda}, where the order-disorder phase transition
finds its counterpart in a chaos-order dynamical transition: this rather
counterintuitive fact can be explained theoretically 
\cite{Firpo} and is likely to be a peculiarity of mean-field models.
We shall come back to this issue in the following, in 
particular in Section \ref{sec_geometry} when we will consider geometric 
properties which are strictly 
related with chaotic dynamics, and which, indeed, exhibit 
a much clearer behaviour near $\varepsilon_c$. 

\section{Geometry of dynamics in the $\varphi^4$ model}

\label{sec_geometry}

Let us turn to the geometrization of the dynamics and its relations with
the dynamical description of the phase transition. We will not enter 
the details recalling only the notations and the main results: all the
details can be found in Ref. \cite{CCP} and in
references quoted therein.

The geometrical formulation of the dynamics of conservative systems
was first used by Krylov in his studies on the dynamical
foundations of statistical mechanics \cite{Krylov} and subsequently
became a standard tool to study abstract systems in ergodic theory.
Several new contributions to this subject appeared in the last years
\cite{others,Pettini,prl95,CCP}. 

Let us briefly recall that the geometrization of the dynamics 
of $N$-degrees-of-freedom systems defined by a Lagrangian
${\cal L} = T - V$, in which the kinetic energy is quadratic in the velocities:
\beq
T=\frac{1}{2}a_{ij} \dot{q}^i\dot{q}^j~,
\eeq
stems from the fact that
the natural motions are the extrema
of the Hamiltonian action functional ${\cal S}_H = 
\int {\cal L} \, dt$, 
or of the Maupertuis' action
${\cal S}_M = 2 \int T\, dt$.
In fact also the geodesics of Riemannian and pseudo-Riemannian 
manifolds are the extrema of a functional: the arc-length 
$\ell=\int ds$, with $ds^2={g_{ij}dq^i dq^j}$, 
hence a suitable choice of the metric tensor allows for the 
identification of the arc-length with either ${\cal S}_H$ or 
${\cal  S}_M$, and of the geodesics with the natural motions of the
dynamical system. Starting from ${\cal  S}_M$ the ``mechanical manifold''
is the accessible configuration space endowed with
the Jacobi metric 
\beq
(g_J)_{ij} = [E - V(\{q\})]\,a_{ij}~,
\label{jacobi_metric}
\eeq 
where $V(q)$ is the potential energy and $E$ is the total energy.
A description of the extrema of Hamilton's 
action ${\cal S}_H$ as geodesics of a ``mechanical manifold'' 
can be obtained using Eisenhart's metric 
\cite{Eisenhart} on an enlarged configuration spacetime 
($\{q^0\equiv t,q^1,\ldots,q^N\}$ 
plus one real coordinate $q^{N+1}$), whose arc-length is
\begin{equation}
ds^2 = -2V(\{ q \}) (dq^0)^2 + a_{ij} dq^i dq^j + 2 dq^0 
dq^{N+1}~.
\label{ds2E}
\end{equation}
The manifold has a Lorentzian structure and the dynamical 
trajectories are those geodesics satisfying the condition
$ds^2 = C dt^2$, where $C$ is a positive constant. 
In the geometrical framework, the (in)stability 
of the trajectories is the (in)stability 
of the geodesics, and it is completely determined by the 
curvature properties of the underlying manifold according to
the Jacobi equation \cite{doCarmo}
\begin{equation}
\frac{D^2 J^i}{ds^2} + R^i_{~jkm}\frac{dq^j}{ds} J^k \frac{dq^m}{ds} = 0~,
\label{eqJ}
\end{equation}
whose solution $J$, usually called Jacobi or geodesic variation field,  
locally measures the distance between nearby geodesics; 
$D/ds$ stands for the covariant derivative
along a geodesic and $R^i_{~jkm}$ are the components of 
the Riemann curvature tensor. 
Using the Eisenhart metric (\ref{ds2E}) the relevant part of the Jacobi equation 
(\ref{eqJ}) is  \cite{Pettini,CCP}
\begin{equation}
\frac{d^2 J^i}{dt^2} + R^i_{~0k0}J^k = 0~,~~~~i=1,\dots ,N
\label{eqdintang}
\end{equation}
where the only non-vanishing components of the curvature tensor are
$R_{0i0j}=\partial^2 V/\partial q_i \partial q_j $. Equation 
(\ref{eqdintang}) is the tangent dynamics equation which is commonly used to
measure Lyapunov exponents in standard Hamiltonian systems. 
Having recognized its geometric origin, in Ref.\ 
\cite{CCP} it has been 
devised a geometric reasoning to derive from  Eq.(\ref{eqdintang})
an {\it effective} scalar stability equation that {\it independently} of the
knowledge of dynamical trajectories provides an average measure of their
degree of instability. This is based on two main assumptions:
$(i)$ that the ambient manifold is {\em almost isotropic}, i.e. 
the components of the curvature tensor --- that for an isotropic manifold
(i.e. of constant curvature) 
are $R_{ijkm}=k_0(g_{ik} g_{jm} - g_{im} g_{jk})$, $k_0=const$ 
--- can be approximated by 
\beq
R_{ijkm} \approx k(t)
(g_{ik} g_{jm} - g_{im} g_{jk})
\eeq
along a generic geodesic $\gamma(t)$; $(ii)$
that in the large $N$ limit the  ``effective curvature''
$k(t)$ can be modeled by a gaussian and $\delta$-correlated stochastic 
process. The mean $k_0$ and variance $\sigma_k$ of $k(t)$ 
are given by the average and the r.m.s. fluctuation of the Ricci curvature 
$k_R = K_R/N$ along a geodesic: 
\begin{mathletters}
\beq
k_0 =  {\langle K_R \rangle}/{N}~;
\eeq   
\beq
\sigma^2_k  =  {\langle (K_R - \langle K_R \rangle)^2 \rangle}/{N}~.
\eeq
\end{mathletters} 
The Ricci curvature along a geodesic is defined as
\beq 
K_R = \frac{1}{v^2} R_{ij} \frac{dq^i}{dt}\frac{dq^j}{dt}
\eeq
where $v^2 = \frac{dq^i}{dt}\frac{dq_i}{dt}$ and 
$R_{ij} = R^k_{~ikj}$ is the Ricci tensor; in the case of Eisenhart metric 
it is 
\beq
K_R\equiv \Delta V = \sum_{i=1}^N \frac{\partial^2 V}{\partial q_i^2}~.
\eeq 
The final result
is the  replacement of Eq.(\ref{eqdintang}) with the aforementioned effective 
stability equation which is independent of the dynamics and is in the
form of a stochastic oscillator equation
\cite{prl95,CCP}
\begin{equation}
\frac{d^2\psi}{dt^2} + k(t) \, \psi = 0~,
\label{eqpsi}
\end{equation}
where $\psi^2 \propto 
|J|^2$. The exponential 
growth rate $\lambda$ of the solutions of Eq. (\ref{eqpsi}),
which is therefore an estimate of the largest Lyapunov
exponent, can be computed exactly: 
\begin{equation}
\lambda = \frac{\Lambda}{2} - \frac{2 k_0}{3 \Lambda}\,,~~
\Lambda = \left(2\sigma_k^2 \tau +
\sqrt{\frac{64 k_0^3}{27} + 4\sigma_k^4 \tau^2}~\right)^\frac{1}{3}
\label{lambda}
\end{equation}
where
$\tau = \pi\sqrt{k_0}/(2\sqrt{k_0(k_0 + \sigma_k)}
+\pi\sigma_k )$;
in the limit $\sigma_k/k_0 \ll 1$ one finds
\beq
\lambda \propto \sigma_k^2~.
\eeq
The latter result is the deep origin of the vanishing of the Lyapunov
exponent in the whole disordered phase of a mean-field rotator
model \cite{RuffoRapisarda}; in fact, in that model, 
the curvature fluctuations can be analytically computed and
turn out to be nonzero in the low-energy region but
vanishing in the whole high-energy phase \cite{Firpo},
consistently with the mean-field character of the model. 
At the critical energy $\sigma_k$ is discontinuous.
The work reported in Ref.\ \cite{Firpo} is particularly important because
it is the first example in which the relationship between chaos
and phase transitions is theoretically investigated within
a framework which allows analytical calculations.
The power of the geometric approach is evident in that case.

It is
natural to wonder whether 
the curvature fluctuations show any remarkable (singular)
behaviour also in correspondence with non mean-field 
phase transitions. Numerical results suggesting a
positive answer to this question have already been found 
in the cases of planar spin models \cite{cccp} and 
three-dimensional $O(n)$ $\varphi^4$ models \cite{Gatto};
in correspondence with a second order phase transition the
curvature fluctuations show a cusp-like behaviour in all 
the cases considered up to now.
These singular behaviours of the curvature fluctuations have been
conjectured to be a consequence of a major change in the global
geometry, if not in the topology, of the mechanical manifolds.

In the following we are going to show that also in 
the case of the two-dimensional $\varphi^4$
lattice theory the above scenario is confirmed. 
Moreover, we will present some results 
concerning other geometric quantities, different from those
defined in the framework of Eisenhart's metric; these results
lend strong support to the topologic intepretation of the
apparently singular behaviour of the curvature fluctuations.

\subsection{Curvature fluctuations with Eisenhart's metric}
\label{sec_eisenhart}
\begin{figure}
\centerline{\psfig{file=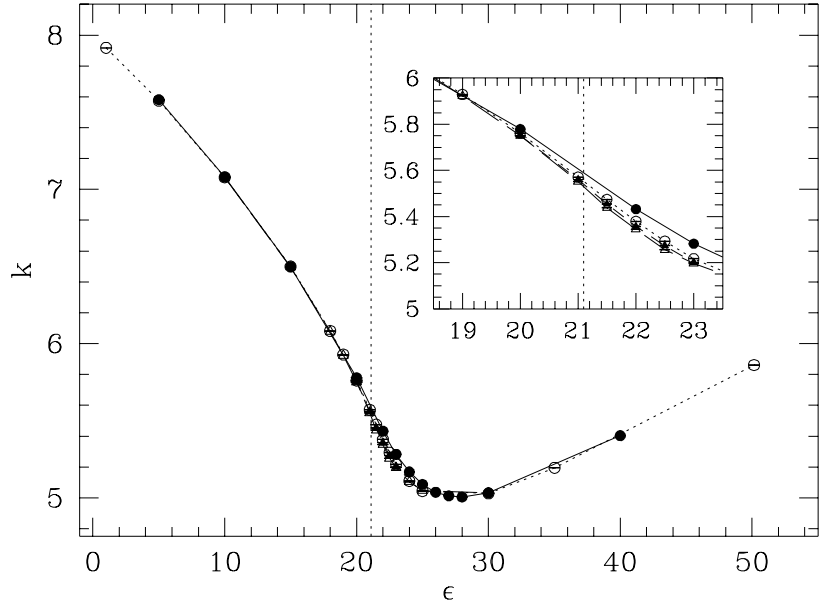,height=9cm,clip=true}}
\caption{Average 
Ricci curvature $k$ (\protect\ref{k_R_phi4}) with Eisenhart 
metric vs. energy density 
$\varepsilon$ for different sizes of the system. The symbols 
denote respectively $N = 10^2$ (solid circles), $N = 20^2$ 
(circles), $N = 30^2$ (solid triangles), $N = 50^2$ (triangles).
The vertical dotted line marks the estimated value of 
$\varepsilon_c$. The inset shows a magnification of the 
transition region.}
\label{fig_k}
\end{figure}

In the case of the lattice $\varphi^4$ theory (\ref{hfi4d}) 
the Ricci curvature per degree of freedom of the Eisenhart
metric is
\beq
k_R = \frac{K_R}{N} = 2dJ - m^2 + \frac{1}{N}  
\sum_{i=1}^N \frac{\lambda}{2}\varphi_{i}^2 ~.
\label{k_R_phi4}
\eeq
The time average $k = \langle k_R \rangle$
of $k_R$ is plotted vs. the energy 
density for the two-dimensional model in Fig. \ref{fig_k}. 
It is evident that 
$k(\varepsilon)$ changes its convexity close to $\varepsilon_c$, 
hence it shares this feature with the temperature.
Anyhow in other models the shape of 
$k(\varepsilon)$ is completely different. 

\begin{figure}
\centerline{\psfig{file=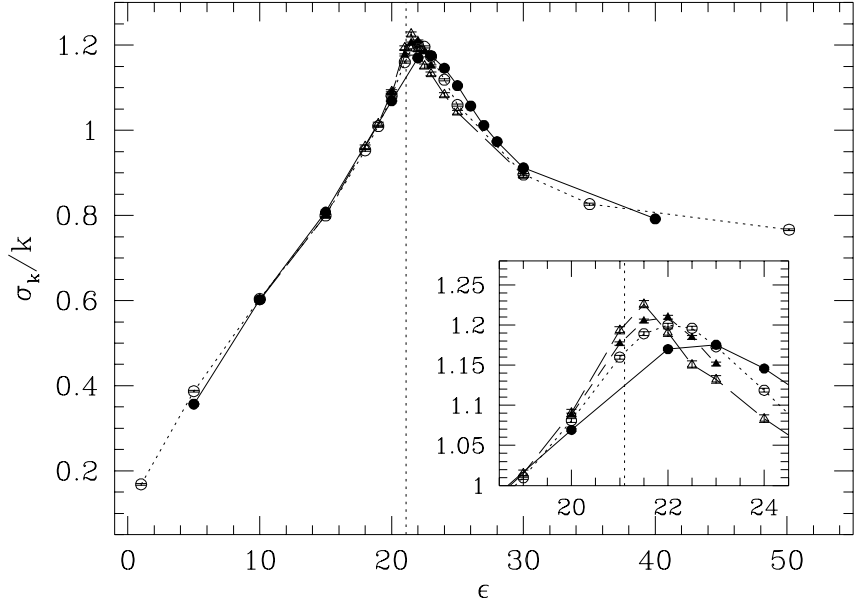,height=9cm,clip=true}}
\caption{Normalized Ricci curvature fluctuations $\sigma_k/k$  
with Eisenhart metric vs. energy density 
$\varepsilon$ for different sizes of the system. The symbols 
are the same as in Fig. \protect\ref{fig_k}.}
\label{fig_flutt_k}
\end{figure}

What is more interesting, and, in the light of the above discussion, 
much more significant, is the behaviour of the 
fluctuations $\sigma^2_k$, reported in Fig. 
\ref{fig_flutt_k}. Actually in this 
figure the normalized fluctuation $\sigma_k/k$ is reported. A cusp-like 
behaviour of the curvature fluctuations is evident
in correspondence 
of the critical energy.

\subsection{Other geometric observables}

\begin{figure}
\centerline{\psfig{file=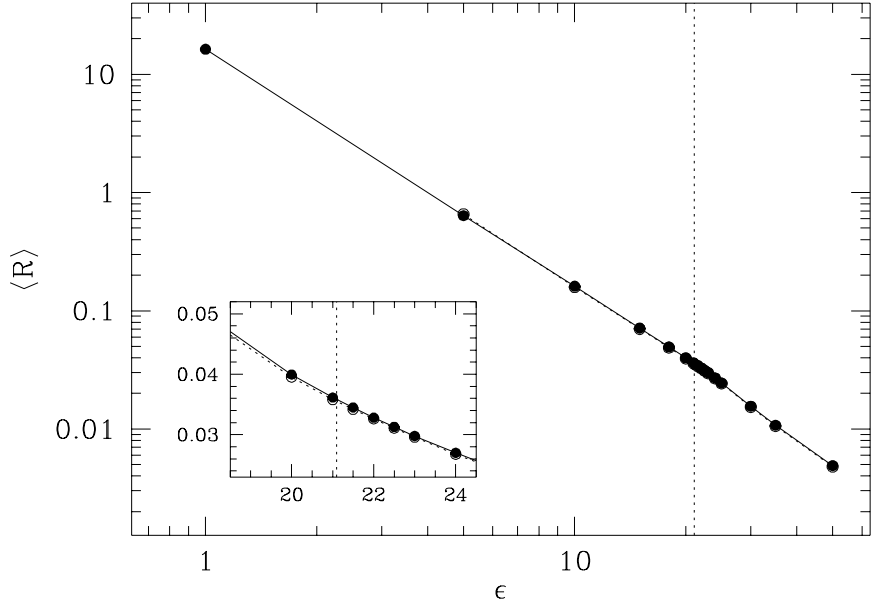,height=9cm,clip=true}}
\caption{Average 
scalar curvature $\langle {\cal R} \rangle$ (\protect\ref{scalar_curv}) 
with Jacobi 
metric vs. energy density 
$\varepsilon$ for different sizes of the system. The symbols 
denote respectively $N = 10^2$ (solid circles) and $N = 20^2$ 
(circles).
The vertical dotted line marks the estimated value of 
$\varepsilon_c$. The inset shows a magnification of the 
transition region.}
\label{fig_scaldyn}
\end{figure}

\begin{figure}
\centerline{\psfig{file=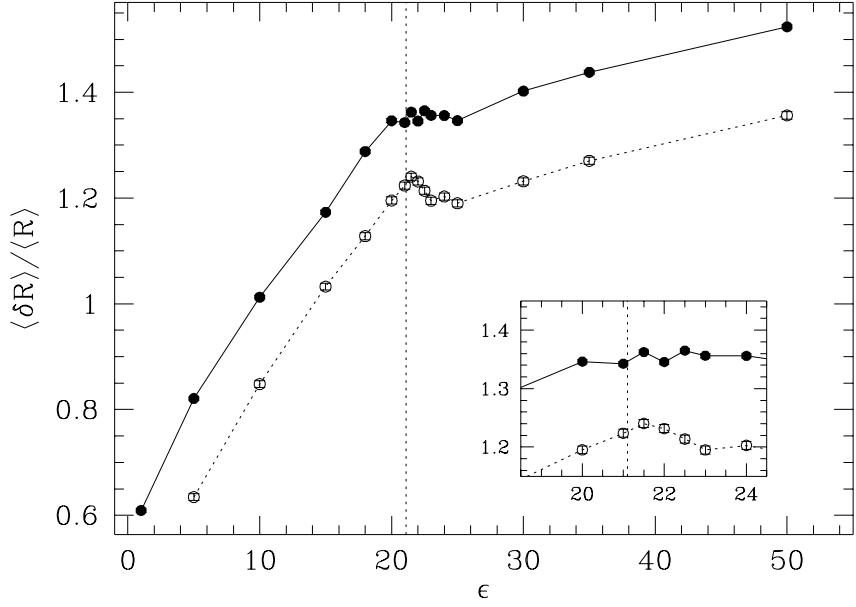,height=9cm,clip=true}}
\caption{Normalized scalar curvature fluctuations 
$\langle\delta{\cal R}\rangle/
\langle {\cal R} \rangle$  
with Jacobi metric vs. energy density 
$\varepsilon$ for different sizes of the system. The symbols 
are the same as in Fig. \protect\ref{fig_scaldyn}.}
\label{fig_fluttdyn}
\end{figure}

The same analysis of the time-averaged geometric quantities can be
carried on also in the framework of the Jacobi metric --- 
Eq.\ (\ref{jacobi_metric}). In this case 
the ambient space is the accessible
configuration space. 
We have studied the behaviour of the time average and
of the fluctuations of the scalar curvature, whose corresponding
dynamical observable is \cite{Pettini,thesis}
\beq
{\cal R} = \frac{N - 1}{4(E-V)^2} \left[ 2(E-V) \triangle V -
(N-6)|\nabla V|^2 \right]~.
\label{scalar_curv}
\eeq
The results are reported in Figs. \ref{fig_scaldyn} and \ref{fig_fluttdyn}.
We see that also with the Jacobi metric the average curvature seems to have a
smooth behaviour near $\varepsilon_c$, even if in the two phases the overall
behaviours of the curvature seem different; on the contrary, the normalized
fluctuation shows a sharp increase near criticality. Notice that these
results have been obtained for small lattices only (up to $20 \times 20$)
hence, in analogy with the other observables, we expect that in larger 
lattices this effect should be even more pronounced. 

Up to now geometry has been introduced through dynamics, by 
identifying the dynamical trajectories with 
the geodesics of suitable manifolds. 

However, other complementary approaches are also possible and interesting.
Let us consider in particular the following one, 
recently introduced \cite{Casartelli} in connection with the 
existence of different regimes in the dynamics of 
natural Hamiltonian systems. Given the dynamics, one can 
study the geometry of the trajectories as curves in the 
phase space, endowed with the Euclidean metric. In this way 
the trajectories are no longer geodesics of any manifold.
Nevertheless, being the Hamiltonian trajectories constrained 
on the constant-energy 
hypersurface $\Sigma_E$, the geometry of such curves 
carries informations on the ``shape'' of the invariant 
hypersurface. Such a 
relation can be made precise in terms of the geometry of 
$\Sigma_E$ seen as a submanifold of ${\bf R}^n$ \cite{thesis}. 
Let us define the curvature of the trajectory 
$x(t) = (\varphi_1(t), \ldots, \varphi_N(t), \pi_1(t), 
\ldots, \pi_N(t))$ in phase space as the generalization to 
$n = 2N$ dimensions of the curvature of a curve in 
the two-dimensional Euclidean plane:
\beq
\kappa = \frac{d\tau}{ds}\,\, ,
\label{kappa}
\eeq
where $\tau$ is the unit tangent vector to the curve $x(t)$, 
i.e. $\tau = \dot x/|x|$, and $s$ is the arc-length induced on 
$x$ by the Euclidean metric of ${\bf R}^n$, hence $ds/dt = 
|{\rm grad \,} {\cal H}| $.

\begin{figure}
\centerline{\psfig{file=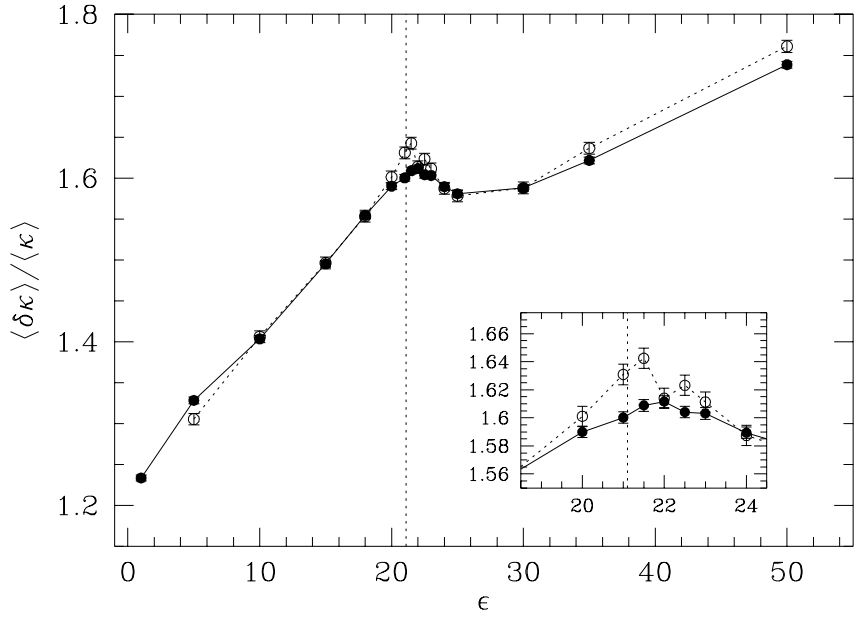,height=9cm,clip=true}}
\caption{Normalized curvature fluctuations $\langle \delta \kappa 
\rangle / \langle \kappa \rangle$ of the trajectories in 
phase space vs. energy density 
$\varepsilon$ for two different sizes of the system: $N =
10^2$ (solid circles) and $N=20^2$ (circles).}
\label{fig_flutt_kappa}
\end{figure}

The time average and the fluctuations of the observable 
$\kappa$ can be defined as usual. In particular, the result 
for the normalized fluctuation $\langle \delta^2 \kappa 
\rangle / \langle \kappa \rangle$ is reported in Fig. 
\ref{fig_flutt_kappa} for the two-dimensional $\varphi^4$ 
model on a square lattice. We observe that the 
fluctuation of this curvature has a cusp-like behaviour in 
correspondence of the transition. Again we stress 
that this ``quasi-singular'' behaviour is obtained already
with extremely small
lattices.

The phenomenology summarized above suggests that in 
correspondence of a phase transition the geometry of the 
manifolds underlying the dynamics undergoes a dramatic 
change. The precise nature of this change is still to be 
understood. Nevertheless, the fact that the singular behaviour
of the curvature fluctations shows up using {\em different} 
geometric settings indicates that it is a 
consequence of some {\em deeper}
property: the topological interpretation of these phenomena \cite{cccp} 
is strongly supported.

\subsection{Geometric observables and Lyapunov exponents}
 
\begin{figure}
\centerline{\psfig{file=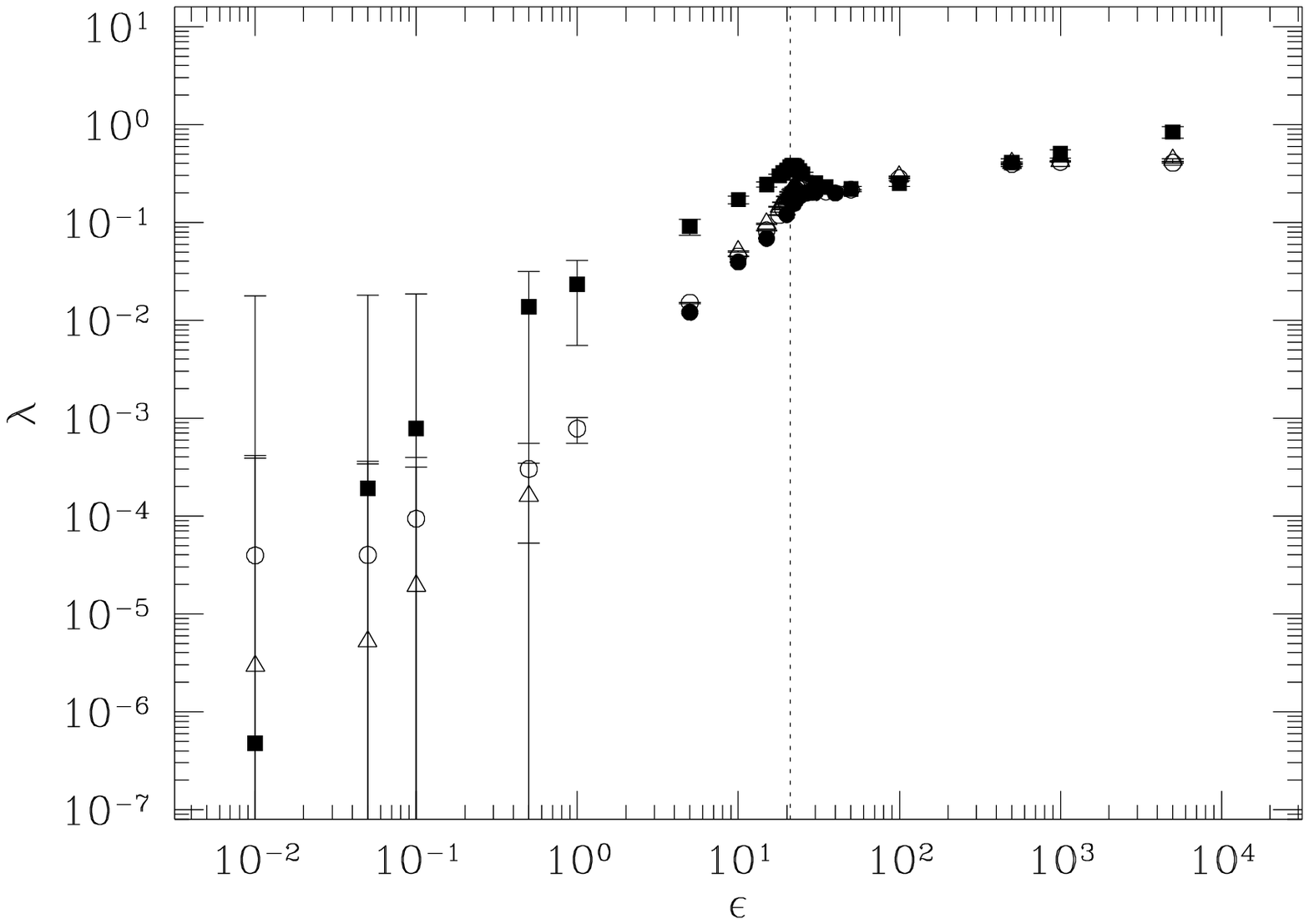,height=9cm,clip=true}}
\caption{Lattice $\varphi^4$ model. Synopsis of the 
geometric prediction of the Lyapunov exponent according to 
Eq. (\protect\ref{lambda}) (full squares) and of the 
numerical simulation results already plotted in Fig. 
\protect\ref{fig_lyap_log}.}
\label{fig_lyap_geo}
\end{figure}

The geometric observables considered in Sec. \ref{sec_eisenhart} 
can be used to estimate the Lyapunov exponents.
The result 
obtained applying Eq.\ (\ref{lambda}) to the 
two-dimensional $\varphi^4$ model are reported in Fig. 
\ref{fig_lyap_geo}.
The agreement between theory and simulations is good qualitatively,
but not from a quantitative point of view. A very 
good agreement is found only in an energy range of about two decades 
just above the transition. We do not have, at present, a deep 
explanation. A tentative one might involve the fact that in 
these models, especially close to the transition, the 
manifolds might be highly anisotropic (as witnessed by 
the growth of the curvature fluctuations) and thus the 
quasi-isotropy assumption might be no longer valid.
Moreover, finite-size effects are expected to play a 
significant role.  
However, the  non-satisfactory agreement between theory and simulation 
for the Lyapunov exponent is not a common feature of all models 
with phase transitions. In fact, in the case of the two- and 
three-dimensional $XY$ models the geometric estimate of the 
Lyapunov exponent is in very good agreement with the 
simulations \cite{cccp}. Moreover, the analytical results obtained
in Ref.\ \cite{Firpo} compare well with simulations. 
In Ref.\ \cite{Gatto} it has also been 
shown that in the case of $\varphi^4$
models the estimate could be improved by adjusting the values of
the timescale $\tau$ in Eq.\ (\ref{lambda}). 

\section{Concluding remarks}

\label{sec_conclusions}

The first result of the present work is that the statistical 
behaviour emerging from the microscopic Hamiltonian dynamics
of a two-dimensional $\varphi^4$ model is perfectly consistent with 
the predictions of equilibrium statistical mechanics, and the 
agreement between the two descriptions includes critical behaviour.
This means that all the tools coming from Hamiltonian mechanics
can be used to investigate phase transitions. One of the aspects
that can be considered in this perspective is certainly the
possibility of an {\em ab initio} description of critical slowing
down. 

A further result is that 
the study of intrinsically dynamical observables, as Lyapunov
exponents, reveals an intriguing relationship between the local
instability properties of the phase-space dynamics and the global
phenomenon of the phase transition. 
In this perspective it is worth mentioning
a recently proposed approach \cite{GozziReuter} that unifies the description of 
Hamiltonian dynamics with the description of its stability through 
a path-integral formalism: this could be a way to relate stability
properties with the phase transition from a general point of view.
Here we approached this problem via a geometrization of the dynamics based on
simple Riemannian tools. We found that the fluctuations of
the curvatures of 
suitably defined manifolds associated with the Hamiltonian
dynamics exhibit a ``singular'' behaviour at the transition.
This fact is coherent with the results recently 
found for other models undergoing a phase transition, 
as the lattice $\varphi^4$ theory in three space dimensions,
with both ${\bf Z}_2$ and $O(n)$ symmetries
\cite{Gatto}, and the classical $XY$ Heisenberg model in two
and three space dimensions \cite{cccp,cecilia_th}. 

A topological conjecture has been proposed to explain this behaviour
\cite{cccp}: the phase transition could be a consequence of
a major topological change in the manifolds where the 
dynamical trajectories live. Such a conjecture receives further 
support from the results presented here, also because an apparently singular
behaviour in the curvature fluctuations is found also using
different geometric settings with respect to those used in
the previous works, which were all based on the Eisenhart 
metric. The problem of a precise characterization of
these topological changes is still open and work is
in progress in this direction (see Ref.\ \cite{thesis} for 
some preliminary results). 

\acknowledgments

We thank S. Caracciolo, C. Clementi, E. G. D. Cohen, M.-C. Firpo,
R. Gatto, R. Livi, M. Modugno, G. Pettini, 
M. Rasetti, and S. Ruffo for fruitful 
discussions and for their interest and support.
Large parts of this work have been done while one of the authors (LC)
was a PhD student at the Scuola Normale Superiore, Pisa, Italy, and 
visiting fellow at the
D\'epartement de Physique Th\'eorique, Universit\'e de Gen\`eve, 
Switzerland. The ISI Foundation in Torino, Italy, is also acknowledged
for its kind hospitality.

\begin{references}

\bibitem[*]{lando} During the editing of the present paper Lando Caiani
tragically died.
\bibitem[\dagger]{lapo} E-mail: lapo@polito.it
\bibitem[\ddagger]{marco} also at INFN, sezione di Firenze, and
INFM, unit\`a di Firenze, Largo Enrico Fermi 2, I-50125 Firenze, Italy.
E-mail: pettini@arcetri.astro.it.
\bibitem{FPU} E. Fermi, J. Pasta, and S. Ulam, 
	Los Alamos Report LA-1940 (1955), in {\it Collected papers of Enrico 
	Fermi}, edited by E. Segr\'e, (University of Chicago, Chicago,
	1965), Vol. 2, p. 978.
\bibitem{others} M. C. Gutzwiller, J. Math. Phys. {\bf 18}, 806 (1977);
J. F. C. van Velsen, J. Phys. A: Math. Gen. {\bf 13}, 833 (1980);
G. K. Savvidy, Nucl. Phys. B {\bf 246}, 302 (1984);
A. Knauf, Comm. Math. Phys. {\bf 110}, 89 (1987).
\bibitem{Pettini} M. Pettini, Phys. Rev. E {\bf 47}, 828 (1993).
\bibitem{CasettiPettini} L. Casetti and M. Pettini, 
Phys. Rev. E {\bf 48}, 2340 (1993);
\bibitem{CerrutiPettini} M. Cerruti-Sola and M. Pettini, Phys. Rev. E 
{\bf 51}, 53 (1995), and Phys. Rev. E {\bf 53}, 179 (1996).
\bibitem{PettiniValdettaro} M. Pettini and R. Valdettaro, Chaos {\bf 5}, 646
(1995).
\bibitem{prl95} L. Casetti, R. Livi, and M. Pettini, Phys. Rev. Lett. {\bf 74},
375 (1995).
\bibitem{CCP} L. Casetti, C. Clementi, and M. Pettini,
Phys. Rev. E {\bf 54}, 5969 (1996).
\bibitem{cccp} L. Caiani, L. Casetti, C. Clementi, and M. Pettini,
Phys. Rev. Lett. {\bf 79}, 4361 (1997).
\bibitem{Ruffo} M. Antoni and S. Ruffo, Phys. Rev. E {\bf 52}, 2361
(1995).
\bibitem{Antoni} Y. Elskens and M. Antoni, Phys. Rev. E {\bf 55}, 6575
(1997).
\bibitem{Rugh} H. H. Rugh, Phys. Rev. Lett. {\bf 78}, 772 (1997).
\bibitem{Giardina} C. Giardin\`a and R. Livi, preprint {\tt
chao-dyn/9709015}, submitted to J. Stat. Phys.
\bibitem{Gross} D. H. E. Gross and M. E. Madjet, 
preprint {\tt cond-mat/9611192}.
\bibitem{Dellago} Ch. Dellago, H. A. Posch, and W. G. Hoover, 
Phys. Rev. E {\bf 53}, 1485 (1996); 
Ch. Dellago and H. A. Posch, Physica A {\bf 230}, 364 (1996).
\bibitem{HohenbergHalperin} P. C. Hohenberg and B. I. Halperin, 
Rev. Mod. Phys. {\bf 49}, 435 (1977).
\bibitem{Goldenfeld} N. Goldenfeld, {\em Lectures on phase transitions
and the renormalisation group} (Addison-Wesley, New York, 1992). 
\bibitem{ButeraCaravati}
P. Butera and G. Caravati, Phys. Rev. A {\bf 36}, 962 (1987).
\bibitem{thesis} L. Casetti, {\em Aspects of dynamics, geometry and
statistical mechanics in Hamiltonian systems}, PhD thesis (Scuola
Normale Superiore, Pisa, 1997) available at the URL
{\tt http://www.sns.it/html/ClasseScienze/ThPhysics.html}.
\bibitem{Gatto} L. Caiani, L. Casetti, C. Clementi, G. Pettini,
M. Pettini, and R. Gatto, preprint {\tt hep-th/9706081}, submitted
to Phys. Rev. E.
\bibitem{algo} L. Casetti, Physica Scripta {\bf 51}, 29 (1995).
\bibitem{Khinchin} A. Ya. Khinchin, {\em Mathematical foundations
of statistical mechanics} (Dover, New York, 1949).
\bibitem{Pearson} E. M. Pearson, T. Halicioglu, and W. A. Tiller, 
Phys. Rev. A {\bf 32}, 3030 (1985).
\bibitem{LPV} J. Lebowitz, J. Percus, and L. Verlet, Phys. Rev. 
{\bf 153}, 250 (1967).
\bibitem{Como} {\em Monte Carlo and Molecular Dynamics of Condensed Matter
Systems}, proceedings of the Euroconference and Summer School held in
Villa Olmo, Como, July 1995, edited by K. Binder and G. Ciccotti (Editrice
Compositori, Bologna, 1996).
\bibitem{Frenkel} D. Frenkel, in Ref. \protect\cite{Como}.
\bibitem{Dunweg} B. Dunweg, in Ref. \protect\cite{Como}.
\bibitem{Binder} K. Binder, Z. Phys. B {\bf 43}, 119 (1981).
\bibitem{Young} A. P. Young, in Ref. \protect\cite{Como}.
\bibitem{Sokal} A. Sokal, {\em Monte Carlo methods: fondations and new
algorithms}, Cours de $3^{\rm eme}$ cycle de la physique en Suisse
Romande, Lausanne, 1989.
\bibitem{Rapisarda} A. Bonasera, V. Latora, A. Rapisarda, Phys. Rev. Lett. 
{\bf 75}, 3434 (1995).
\bibitem{Duke_pre} C. S. O'Hern, D. A. Egolf, H. S. Greenside, 
Phys. Rev. E {\bf 53}, 3374 (1996).
\bibitem{Mehra} V. Mehra, R. Ramaswamy, preprint {\tt chao-dyn/9706011}.
\bibitem{RuffoRapisarda} V. Latora, A. Rapisarda, and S. Ruffo,
{\tt chao-dyn/9707024}, submitted to Phys. Rev. Lett.
\bibitem{Firpo} M.-C. Firpo, preprint (1997).
\bibitem{Krylov} N. S. Krylov, {\em Works on the foundations of 
statistical mechanics} (Princeton University Press, Princeton, NJ, 1979).
\bibitem{Eisenhart} L. P. Eisenhart, Ann. Math. {\bf 30}, 591 (1929).
\bibitem{doCarmo} M. P. do Carmo, {\em Riemannian geometry}
(Birkh\"auser, Boston-Basel, 1993).
\bibitem{VanKampen} N. G. Van Kampen, Phys. Rep. {\bf 24}, 71 (1976). 
\bibitem{Casartelli} C. Alabiso, N. Besagni, and M. Casartelli, 
J. Phys. A: Math. Gen. {\bf 29}, 3733 (1996).
\bibitem{GozziReuter} E. Gozzi and M. Reuter, Phys. Lett. B {\bf 233}, 
383 (1989); E. Gozzi, M. Reuter and W.D. Thacker, Chaos, Solitons \&
Fractals {\bf 2}, 441 (1992); E. Gozzi and M. Reuter, {\em ibid.} 
{\bf 4}, 1117 (1994).
\bibitem{cecilia_th} C. Clementi, Laurea thesis, University of Florence,
Italy (1995).
\end{references}
\end{document}